# The JPEG Pleno Learning-based Point Cloud Coding Standard: Serving Man and Machine

**ANDRÉ F. R. GUARDA[1], (Member, IEEE), NUNO M. M. RODRIGUES[1,2], (Senior Member, IEEE), AND FERNANDO PEREIRA[1,3], (Fellow, IEEE)**

[1]Instituto de Telecomunicações, 1049-001 Lisbon, Portugal
[2]ESTG, Politécnico de Leiria, 2411-901 Leiria, Portugal
[3]Instituto Superior Técnico, Universidade de Lisboa, 1049-001 Lisbon, Portugal

Corresponding author: André F. R. Guarda (e-mail: andre.guarda@lx.it.pt).

This work was supported by Fundação para a Ciência e Tecnologia, I.P. (FCT, Funder ID = 50110000187) by project reference UIDB/50008/2020 with DOI identifier 10.54499/UIDB/50008/2020, and project PTDC/EEI-COM/1125/2021 entitled "Deep Learning-based Point Cloud Representation."

**ABSTRACT** Efficient point cloud coding has become increasingly critical for multiple applications such as virtual reality, autonomous driving, and digital twin systems, where rich and interactive 3D data representations may functionally make the difference. Deep learning has emerged as a powerful tool in this domain, offering advanced techniques for compressing point clouds more efficiently than conventional coding methods while also allowing effective computer vision tasks performed in the compressed domain thus, for the first time, making available a common compressed visual representation effective for both man and machine. Taking advantage of this potential, JPEG has recently finalized the JPEG Pleno Learning-based Point Cloud Coding (PCC) standard offering efficient lossy coding of static point clouds, targeting both human visualization and machine processing by leveraging deep learning models for geometry and color coding. The geometry is processed directly in its original 3D form using sparse convolutional neural networks, while the color data is projected onto 2D images and encoded using the also learning-based JPEG AI standard. The goal of this paper is to provide a complete technical description of the JPEG PCC standard, along with a thorough benchmarking of its performance against the state-of-the-art, while highlighting its main strengths and weaknesses. In terms of compression performance, JPEG PCC outperforms the conventional MPEG PCC standards, especially in geometry coding, achieving significant rate reductions. Color compression performance is less competitive but this is overcome by the power of a full learning-based coding framework for both geometry and color and the associated effective compressed domain processing.

## I. INTRODUCTION

3D models provide a rich and immersive way to represent visual data associated to objects, scenes, and environments, greatly enhancing user experiences across numerous applications, ranging from virtual/augmented/mixed reality for entertainment, to autonomous driving, or even architecture and engineering. By adopting 3D models, applications can offer 6-Degrees of Freedom experiences that improve perception of depth, scale, perspective, and freedom of navigation, making them far more interactive and realistic than traditional 2D images. The continuous advancement of acquisition, processing, and rendering technologies has broadened the scope and accessibility of 3D models, making them indispensable tools in modern visualization and communication [1].

Achieving rich 3D visual representations requires describing the complete information about the light rays in a scene, namely the intensity of the light at each position $(x, y, z)$, for any viewing angle $(\theta, \phi)$, over time $(t)$ and for any wavelength $(\lambda)$. This led to the conceptualization of the 7D mathematical definition of the so-called *plenoptic function* $P(x, y, z, \theta, \phi, t, \lambda)$ [2]. New imaging modalities such as light fields, point clouds (PCs), and meshes, have appeared, attempting to approximate the plenoptic function to provide such 3D visual representations. Among these, PCs are gaining relevance due to the growing availability of acquisition sensors, their versatility and easy interactivity, and their lightweight nature, which supports real-time processing. A PC consists of an unordered set of points in 3D space representing the surface of an object or scene, which may be static or dynamic, i.e., varying in time. The (*x*, *y*, *z*) coordinates define the geometry of the PC, which may also contain attributes such as color, reflectance, or normal vectors, among others. Naturally, these richer imaging modalities present unique characteristics that pose new challenges for the development of efficient coding techniques.

To address these challenges, the JPEG Pleno standard was launched in 2015, aiming to design a standard framework for representing and exchanging plenoptic imaging modalities [3]. Light fields were the first focus, with a Call for Proposals (CfP) on Light Field Coding in January 2017 [4]. Later, in July 2020, JPEG issued a Call for Evidence (CfE) on Point Cloud Coding (PCC) [5], addressing in particular "coding technologies for static point cloud content that enable scalable decoding of the bitstream and random access to subsets of the point cloud" [5]. Since, at that time, the Geometry-based Point Cloud Compression (G-PCC) and Video-based Point Cloud Compression (V-PCC) MPEG standards [6][7] were already well under way, JPEG decided to focus on scalability and random access requirements to differentiate itself. The outcome was somehow unexpected since the single submitted proposal was a deep learning (DL)-based PC coding solution – a rather new technological approach in multimedia coding standardization bodies at that time.

By 2020, DL-based coding for multimedia data, particularly images, was gaining momentum due to the competitive compression performance [8][9][10] and the recent history of success by DL-based solutions in computer vision tasks , such as classification, recognition and detection, in comparison to the traditional and conventional hand-crafted counterparts. At the same time, with more and more multimedia content being consumed by machines, designing coding solutions specifically targeting machine consumption became an important research topic [11][12][13][14][15][16]. Traditional codecs targeting human visualization rely on exploiting the human visual system for increased compression efficiency. However, when targeting machine vision applications, using traditional codecs may be undesirable [17][18] since maintaining fidelity at the pixel level does not guarantee that high-level semantics are preserved, potentially hurting the task's performance, and in turn the full-resolution image contains plenty of redundant information for machine vision tasks, making it much less efficient in terms of coding rate.

Ultimately, these developments led JPEG to launch the so-called JPEG AI (from Artificial Intelligence) standard [19] with a CfE [20] and a CfP on Learning-based Image Coding Technologies [21] launched in January 2020 and January 2022, respectively. The scope of JPEG AI was defined as the creation of a learning-based image coding standard offering an efficient compressed representation, targeting both human visualization as well as an effective performance for image processing and computer vision tasks [21]. This marked a significant conceptual advancement in image coding, as it was the first time a unified image representation language was designed for both human visualization and machine consumption. From the CfP results [22], issued in July 2022, JPEG AI reports average rate reductions above 30% over the Versatile Video Coding (VVC) standard [23] in its Intra coding mode, the most efficient conventional image coding solution available. These are impressive compression performance gains, especially considering that the VVC Intra codec is the result of decades of worldwide research efforts in hand-crafted image codecs. Moreover, significant performance and complexity benefits were shown in terms of image classification when the same coded streams used for human visualization were also directly used for DL-based compressed domain classification, in comparison with the usual decompressed domain classification where the images are first (lossy) decoded and only after classified [22].

Following these developments, in January 2022, JPEG Pleno PCC launched a CfP [24] shifting its scope to address "learning-based coding technologies for static PC content and associated attributes with emphasis on both human visualization and decompressed/reconstructed domain 3D processing and computer vision with competitive compression efficiency compared to PC coding standards in common use, with the goal of supporting a royalty-free baseline," targeting use cases such as "computer-aided manufacturing, entertainment, virtual and augmented reality display, cultural heritage preservation, autonomous navigation and remote sensing and geographical information systems."; this scope is aligned with JPEG AI's, showing the JPEG intent to start building a set of learning-based coding standards for several imaging modalities. Fig. 1 shows the learning-based coding framework targeted by JPEG Pleno PCC (and JPEG AI) which serves three key purposes, to be addressed in three sequential stages: in stage 1, standard decoding/reconstruction for human visualization (this stage was the target of the January 2022 CfP); in stage 2, additionally supporting compressed domain processing for enhanced human visualization; and in stage 3, additionally supporting compressed domain computer vision tasks, always from the same coded stream. In summary, in addition to targeting a competitive rate-distortion (RD) performance, the JPEG Pleno PCC standard aims to provide a unified compressed representation approach for both man and machine, enabled by the usage of DL, which offers two key advantages: first, it reduces the computational complexity of machine tasks by bypassing decoding and feature extraction, using instead the compressed domain latent features already available; and second, it improves task performance, particularly at lower rates where compression artifacts can degrade feature extraction, since features were obtained from the original data.

This paper reports in first-hand the newly finalized JPEG PCC standard (the authors of this paper have been main contributors to the standard) to the multimedia coding research community, as its development details are not available to the general public; it does not intend to propose a novel solution. The main contributions are as follows:

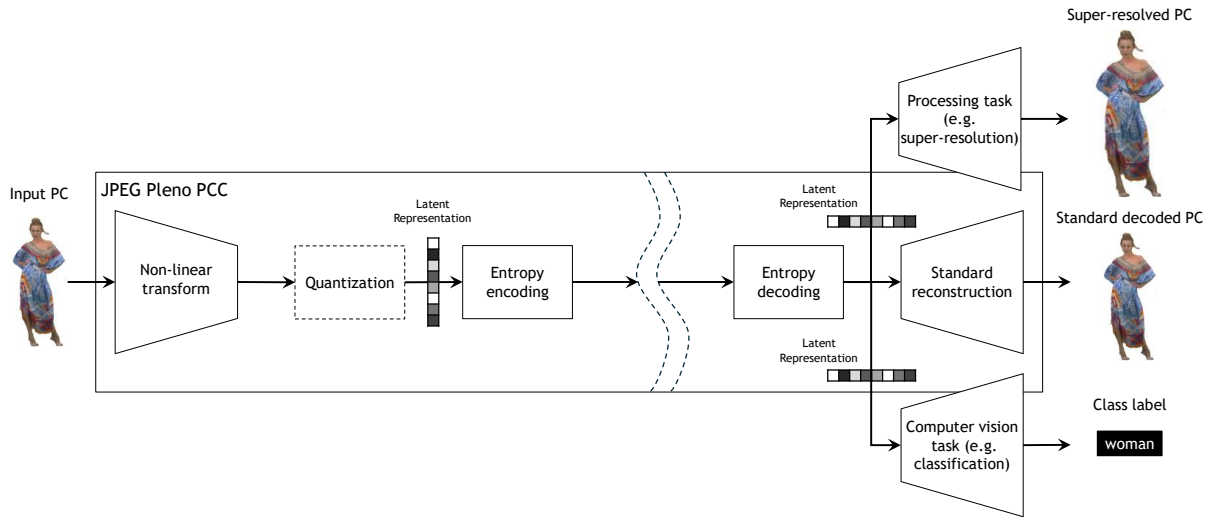

**FIGURE 1.** JPEG Pleno Learning-based PCC framework for both man and machine consumption.

- Offering a detailed overview of the recent JPEG Pleno Learning-based Point Cloud Coding standard (from here on labelled JPEG PCC).
- Performing a comprehensive performance assessment of JPEG PCC in comparison to state-of-the-art PC coding solutions, both conventional and learning-based, using well defined and meaningful test conditions.
- Analyzing the JPEG PCC's main strengths and weaknesses, providing possible directions for future developments inside JPEG and outside in the PC coding research community at large.

The JPEG PCC standard considers the geometry and color attribute, coded sequentially in two dependent pipelines: first the geometry is (lossy) coded independently, and then the color is coded taking into account the already decoded geometry. For geometry coding, the main adopted techniques involve processing the geometry data directly in its original 3D structure, using two DL models for coding and super-resolution (SR), consisting of 3D convolutional neural networks (CNN) with sparse convolutions. On the other hand, for color coding, the main adopted techniques involve projecting the originally 3D color data onto 2D images to be coded with the JPEG AI standard, selected for three main reasons: i) having very high compression efficiency; ii) being a learning-based codec, with the potential to produce a unified compressed domain color representation for both man and machine; and iii) being a JPEG standard, creating a synergy between JPEG learning-based coding standards for different modalities with similar scopes. It is worth noting that while JPEG PCC has only considered a single PC attribute, namely the color/texture, more attributes, e.g., normal vectors or reflectance, may be added in the future.

As for conventional coding standards, the JPEG PCC standard only normatively specifies the decoder behavior as well as the bitstream syntax and semantics; consequently, only the decoder parts of the DL models are normative, including the architecture and the learned parameters themselves. This means that the encoder description presented in Section III is merely informative, based on the JPEG PCC reference software (final evolution of the Verification Model (VM) software used during the standard development). As such, freedom is granted to retrain or redefine the encoder parts of the DL models, thus allowing to optimize the target performance, naturally constrained by the normative bitstream syntax and semantics, and decoder models; this guarantees the flexible creation of JPEG PCC compliant bitstreams with increasing performance, if further complexity is invested.

In terms of RD performance, JPEG PCC performs considerably well for geometry-only coding, outperforming both the conventional MPEG PCC standards for most solid, dense and sparse PCs. Compared to other DL-based PC coding solutions in the literature [25][26][27], JPEG PCC is also the most consistent, achieving a good RD performance for all types of PC content, although not necessarily the best for all PCs. For color coding, the JPEG PCC's RD performance is less competitive, e.g., failing to reach V-PCC Intra for the solid PCs and G-PCC for the sparse PCs. In terms of complexity, for geometry coding, the JPEG PCC standard uses a DL coding model with 5.1 million parameters and a DL SR model with 7.3 million parameters; however, as for most CNN-based models, they are parallelization friendly and can be efficiently run in GPU devices. It is worth noting that this is the first version of the JPEG PCC standard which will be improved in further versions, both in terms of performance and functionality, benefiting from a fast-moving technology. Furthermore, JPEG PCC aims for more than just compression efficiency.

The remainder of the paper is structured as follows: Section II briefly reviews the most relevant literature on conventional and learning-based PC coding. Section III describes in detail the JPEG PCC framework and most important modules. Section IV assesses the JPEG PCC performance and Section V shows the impact of some of the adopted modules by performing ablation and replacement studies. Finally, Section

VI concludes the paper.

A repository with the JPEG PCC decoded PCs from the test dataset and the full experimental results will be made publicly available (https://github.com/aguarda/JPEG-Pleno-PCC-Benchmarking), to enable the research community to use JPEG PCC as a benchmark for new PC codecs.

## II. LITERATURE REVIEW
This section provides a brief review of the state-of-the-art in PC coding, covering both conventional and learning-based approaches, with an emphasis on the latest advancements and research trends.

### A. CONVENTIONAL PC CODING
The two MPEG PCC standards [6][7] represent the leading technology of conventional PC coding, with each one addressing PCs with distinct characteristics using different conceptual approaches. The G-PCC standard was designed for static PCs and leverages the 3D geometry information directly, making it particularly effective for sparse PC content. For geometry coding, G-PCC specifies an Octree-based coding mode based on an octree decomposition of the 3D space, where occupied octree nodes are represented by '1' and empty nodes are represented by '0', resulting in a binary pattern that is entropy coded. In addition to the octree decomposition, G-PCC employs several coding tools to improve the RD performance, such as the direct coding mode and the planar mode, both useful for sparse PCs, or the triangle-soup (Trisoup) coding mode, which terminates the octree decomposition earlier and estimates the points' positions within each voxel using triangles, making it useful for denser PCs. The color attributes are encoded on top of the decoded geometry, after a recoloring process, using either the Region-Adaptive Hierarchical Transform (RAHT) or a Predicting and Lifting (PredLift) transform scheme [7]. The RAHT is a hierarchical spatial transform similar to wavelet in image coding, whereas the PredLift transform is based on a Level of Detail (LOD) generation process, where attributes of each point are predicted from already encoded neighboring points at the same or a lower LOD layer.

In contrast, the V-PCC standard primarily targets dynamic PCs, and it has been optimized for dense surface-like PCs. V-PCC projects the 3D PC data (both geometry and attributes) into 2D image sequences, which are then efficiently coded using established video coding standards, such as High Efficiency Video Coding (HEVC) and VVC. V-PCC initially segments the PC into clusters determined using the normal vectors at each point, followed by a refinement process to smooth the clusters and avoid isolated points. From these clusters, contiguous 3D patches are extracted and then orthogonally projected onto 2D planes along the PC bounding box. The patches are then packed onto multiple 2D images, with the PC geometry information being represented similarly to depth maps, where each pixel denotes the distance to the projection plane, and the PC color information being represented as regular images. Additionally, a binary occupancy map is produced to signal which pixels in the 2D images correspond to 3D points, as well as information of each patch's position to allow the PC reconstruction at the decoder. Finally, the produced 2D images are coded with a 2D image or video codec.

### B. DEEP LEARNING-BASED PC CODING
The success of DL-based image coding has inspired significant interest in applying similar techniques to PC data, with geometry coding being the main focus. PCs are unstructured by definition, which poses a challenge when trying to extract local features and exploit the neighboring information, what is widely done using CNNs. As a result, most works in the literature adopt a volumetric representation, defining a regular 3D grid of voxels and making it possible to apply CNNs. Quach *et al.* [28] proposed one of the first such works, using a simple six-layer autoencoder, quantization, and entropy coding, trained end-to-end using a RD loss function. Quach *et al.* [29] later made several improvements to their work, including: i) added a PC block partitioning scheme with each block being independently coded; ii) developed a deeper and more complex autoencoder architecture including residual blocks; iii) adopted a variational hyperprior for an adaptive entropy coding as proposed by Ballé *et al.* for image coding [9]; and iv) introduced a sequential training strategy to reduce training time.

Wang *et al.* [30] proposed another of the first works, based on an autoencoder architecture with inception-residual blocks [31], quantization, and adaptive entropy coding via a variational hyperprior. In addition to a block partitioning scheme, this work also introduced down-scaling as a pre-processing step to reach lower coding rates, as well as a top-k binarization approach at the decoder rather than a simple thresholding. Wang *et al.* [25] later extended their previous work, introducing a sparse tensor representation and sparse convolutions, which considerably reduce the computational complexity with respect to the regular volumetric representation. The proposed PC geometry codec, labeled PCGCv2, provides a multiscale hierarchical reconstruction of the PC at the decoder, where the PC is progressively reconstructed at different scales by performing voxel classification. For this purpose, the distortion component of the training loss function considers the reconstruction at every scale. Wang *et al.* [32] further improved their work based on the multiscale hierarchical reconstruction idea. The proposed codec, labeled SparsePCGC, is modular in the sense that the DL model is applied between two consecutive scales, and shared across all scales. At each scale, the encoder performs down-sampling, whereas the decoder performs up-sampling and predicts the probability of each sub-voxel being occupied via cross-scale and same-scale context modeling. SparsePCGC is capable of both lossy and lossless coding. For lossless coding, the predicted occupancy probabilities are used for entropy coding the ground truth coordinates. For lossy coding, only the lowest scales are coded losslessly, while the remaining scales perform voxel classification based on the predicted occupancy probabilities. Additionally, SparsePCGC uses slightly different DL models and tools to more efficiently code dense PCs and sparse PCs.

Liu *et al.* [26] proposed to extend PCGCv2 by adding k-nearest neighbors (k-NN)-based self-attention. The proposed codec, labeled PCGFormer, uses the k-NN search to determine the local neighborhood of each point, and applies self-attention on this local neighborhood to better characterize the spatial relations. Lazzarotto *et al.* [33] proposed a two-layer scalable PC geometry coding approach based on residual coding. Considering any existing codec as a base layer, the proposed residual coding neural network receives as input the original PC concatenated with the distorted PC from the base layer. This residual coding neural network consists of an autoencoder, quantization, and a variational hyperprior for adaptive entropy coding architecture, similar to [29], but the decoder additionally includes a CNN-based refiner module that receives the output of the autoencoder and the distorted PC from the base layer, producing the enhancement layer PC. Frank *et al.* [34] later proposed a latent space slicing approach to improve the entropy coding in the commonly adopted autoencoder with variational hyperprior architecture. In the proposed approach, the latent representation produced by the autoencoder is sliced along the channel dimension, with each slice being separately entropy coded. The entropy coding parameters for each slice are estimated using a learned CNN-based transform, receiving as input the global entropy coding parameters produced by the variational hyperprior, as well as the previously decoded slices. Additionally, another learned CNN-based transform is used to reduce the quantization error by predicting the latent residual.

Pang *et al.* [27] proposed a scalable PC geometry codec, labeled GRASP-Net, in which an octree coder is first used to produce a base layer, consisting of a coarse version of the PC, and two neural networks are used to produce a enhancement layer. For the enhancement layer, a geometric subtraction between the input PC and the coarse base layer is first performed to generate a residual point set, containing the residual information of the k-NN for each base layer point. Then, the first neural network, which is point-based, is tasked with extracting features from the residual point set, while the second neural network, which is a sparse CNN, is tasked with producing a latent representation with lower dimensionality, which is then entropy coded. Pang *et al.* [35] later extended GRASP-Net, making the base layer even coarser and adding a third neural network in between the point-based and the sparse CNN to process the finer details. This third neural network is a sparse CNN similar to PCGCv2 [25], enhanced on the decoder side using context modeling for a better voxel classification, and a transformer for a better feature aggregation that captures long-range dependencies. Ahn *et al.* [36] also proposed an improvement over GRASP-Net adding a third neural network in between the point-based and the sparse CNN, with the goal of forcing the probability distributions of the features to be Gaussian, thus creating more descriptive features.

While there has been plenty of focus on PC geometry coding, learning-based PC attributes/color coding is relatively unexplored in the literature, despite being an important aspect of PCs. Alexiou *et al.* [37] proposed an approach for the joint coding of both geometry and color, using a simple autoencoder architecture based on [28]. In this approach, the PC color information is concatenated with the PC geometry information in a volumetric representation with one channel for the geometry (occupancy) and three channels for the color (RGB). The training RD loss function is also adapted to include distortion terms for geometry and color. Quach *et al.* [38] proposed a PC attribute coding solution using a learning-based folding operation. Assuming an already decoded PC geometry, the proposed solution uses a neural network that learns a function for folding a 2D grid onto the decoded PC geometry, then maps the PC attributes onto the 2D grid, which can be coded with a regular 2D image codec. Later, Wang *et al.* [39] extended their previous work on PC geometry coding [25] for PC attribute coding, simply by adding the color attributes as features in the sparse tensor representation of the geometry.

More recently, Guarda *et al.* [40] proposed a joint PC geometry and color coding approach, similar to [37], using an autoencoder with a variational hyperprior architecture. Additionally, the proposed solution includes a down-sampling and a learning-based super-resolution module, used at the encoder and decoder, respectively, with the goal of improving the lower rates RD performance. This solution was submitted as a response to the JPEG Pleno PCC CfP [24], and was selected as the winning proposal, serving as the initial basis for the development of the current JPEG PCC standard described in this paper.

## III. JPEG PCC FRAMEWORK AND MODULES

This section presents in detail the JPEG Pleno Learning-based PC Coding (JPEG PCC) standard for lossy coding of static PCs; the version here presented has reached Final Draft International Standard (FDIS) status in January 2025. After a walkthrough of the encoding and decoding processes, a detailed description of the most important modules is presented. It is worth repeating that the encoding process description is only informative, corresponding to the JPEG PCC reference software, as only the bitstream syntax and semantics and the decoding process are normative.

### A. WALKTHROUGH

The overall JPEG PCC architecture is shown in Fig. 2. The JPEG PCC codec processes the PC geometry and color components in two pipelines, producing two separable bitstreams: the first pipeline encodes/decodes the PC geometry, independently of the color; the second one encodes/decodes the PC color (when available) depending on the (decoded) geometry. Naturally, the color is always dependent on the geometry since a PC cannot exist without geometry but may exist without color. The use of two sequential pipelines generating two separate bitstreams has two main advantages: it allows for one-way component scalability, i.e., decoding only the geometry from the geometry bitstream or the geometry and color from the two bitstreams; and also a more flexible rate/quality control, since geometry and color rate allocation can be defined independently at coding time.

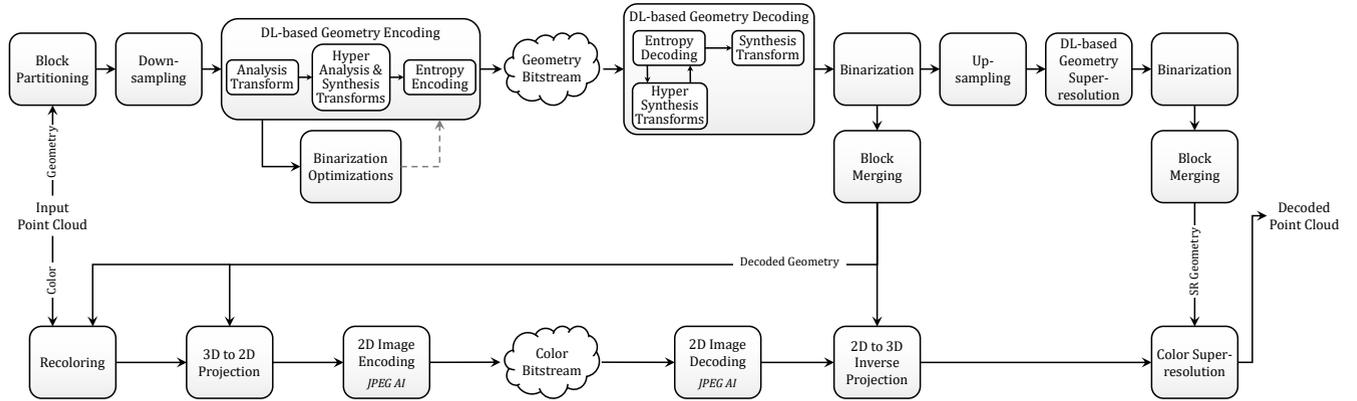

**FIGURE 2.** Overall architecture of the JPEG Pleno Learning-based PC Coding standard, including the PC geometry coding pipeline, shown at the top, and the PC color coding pipeline, shown at the bottom.

The JPEG PCC codec follows a 3D/2D approach, where the geometry coding pipeline processes the PC geometry data directly in its native 3D form, while the color coding pipeline converts/projects the PC data onto a 2D representation to be coded with an image codec. Both pipelines use a DL coding model as the core module. The (informative) PC geometry encoding process works as follows:

- **Block Partitioning**: The input PC geometry, which is assumed to be voxelized, i.e., with integer coordinates, is divided into non-overlapping 3D blocks, with a user-defined size. This allows to independently code each block, providing spatial random access, where the block size controls its granularity. The position of each block in the overall PC is included in the bitstream, to allow for a correct PC reconstruction. Naturally, the full PC may be coded as a single block, depending on the available computational resources (notably memory).

- **Down-sampling**: Each block may be optionally down-sampled to a lower precision using a user-defined sampling factor (SF), which is included in the bitstream, resulting in a lower spatial resolution. By reducing the precision, this down-sampling module reduces the number of PC points as many collapse onto each other, ultimately reducing the amount of information to encode. Furthermore, the geometry is densified as the empty space between points is also reduced. Thus, higher values of SF are generally used when targeting lower rates or/and when encoding sparser PCs.

- **DL-based Geometry Encoding**: Each block is encoded with a voxel-based end-to-end DL coding model, consisting of two sub-networks: an autoencoder, which includes an analysis transform that produces a latent representation, and a variational autoencoder-based hyper network, which performs latent prediction and facilitates the residue entropy encoding. It works as follows:
  - **Analysis Transform**: The encoder part of the autoencoder model applies a learned, non-linear transform, which generates a set of coefficients, referred to as the latent representation.
  - **Hyper Analysis & Synthesis Transforms**: The hyper network has three parts: a hyper analysis transform that produces a hyper latent representation, and two hyper synthesis transforms. While these two both receive the hyper latent representation as input, the first generates a prediction for the latent representation, resulting in a residue to be coded, and the second estimates the residue's probability distribution.
  - **Entropy Coding**: The latents' residue is quantized using a user-defined, real-valued, quantization step (QS), included in the bitstream, and finally entropy coded using the estimated residue's probability distribution. Since it is needed for entropy decoding, the hyper latent representation is also entropy coded, using a fixed probability distribution instead. The entropy coded latents' residue and hyper latent representation constitute the geometry bitstream.

- **Binarization Optimizations**: The output of the DL coding model at the decoder is the probability of each voxel being occupied, i.e., a value between 0 and 1. In order to be able to reconstruct the PC geometry, a binarization method is necessary to select which voxels should be marked as occupied. JPEG PCC adopts a so-called *Top-k* binarization approach performed at the decoder (therefore normative), where the $k_C$ voxels with highest probabilities (as generated by the DL coding model) are selected as occupied. The target value $k_C$ is non-normatively selected at the encoder and included in the bitstream. A possible approach, adopted in the reference software, selects $k_C$ as the value maximizing the reconstruction quality according to a chosen quality metric. Likewise, the output of the DL SR model at the decoder, described later in Section III.C, also requires binarization, for which the same optimization method is adopted in the reference software, selecting a value $k_S$ included in the bitstream.

The (normative) PC geometry decoding process works as follows:

- **DL-based Geometry Decoding**: The geometry bitstream is decoded using the decoder part of the previously mentioned DL coding model, reconstructing each PC geometry block as follows:
  - **Entropy Decoding**: The hyper latent representation is first entropy decoded from the geometry bitstream, since it will be used to generate the probability distribution necessary to afterwards entropy decode the residue.
  - **Hyper Synthesis Transforms**: As performed at the encoder, the hyper synthesis transforms generate both a prediction of the latent representation and the residue's probability distribution. The decoded residue is then added to the prediction to form the latent representation.
  - **Synthesis Transform**: The decoder part of the autoencoder applies the normative inverse transform to the latent representation, to reconstruct the input PC block.
- **Binarization:** The output of the DL coding model is binarized using the Top-k approach, using the $k_C$ value defined at the encoder and included in the bitstream
- **Up-sampling**: Each decoded block is up-sampled back to the original precision, using the same SF used at the encoder for down-sampling and included in the bitstream; the use of this optional module is decided at the encoder, namely by selecting SF > 1. This simple up-sampling approach does not change the number of decoded points, as it simply rescales their coordinates by SF.
- **DL-based Geometry Super-resolution**: After up-sampling, a DL-based post-processing geometry SR module [41] may be applied to each block. While its usage is optional and decided at the encoder, SR can only be applied if SF > 1, thus after the up-sampling module above is applied. The objective of this SR module is to increase the density of the up-sampled block to recover from the down-sampling applied at the encoder. When used, the SR module is able to improve the quality of the decoded PC at no rate cost, especially for originally dense PCs more efficiently coded at lower rates using down-sampling.

- **Binarization**: Similarly to the DL coding model, the output of the DL SR model is also binarized using the Top-k approach, using the $k_S$ value defined at the encoder and included in the bitstream.
- **Block Merging**: The decoded blocks are merged, using the original block positions included in the bitstream, to reconstruct the full decoded PC geometry.

The PC geometry encoding process is lossy for multiple reasons, notably: i) the down-sampling module is likely to remove neighboring points due to the precision reduction which may result in multiple points collapsing on the same position; ii) even if the PC is sparse enough that there are no points collapsing when down-sampling, the up-sampling module does not restore the exact position of each point when the precision is increased; iii) the quantization in the DL-based geometry encoding module inherently introduces a quantization error; and iv) the analysis transform in the DL-based geometry encoding module is irreversible, and thus the synthesis transform cannot mathematically reconstruct the input, even without quantization.

The PC color encoding process, shown in the bottom part of Fig. 2, works as follows:

- **Recoloring**: The (lossy) decoded PC geometry is recolored with the original PC color by copying the color of collocated points and interpolating the color of the remaining points. This allows to encode only the PC color information associated with the PC points reconstructed by the decoder.
- **3D to 2D Projection**: The recolored PC is segmented into 3D patches which are projected onto 2D planes and packed into two images, effectively transforming the 3D color coding problem into a 2D image coding problem.
- **2D Image Encoding**: The two 2D images with the color information are encoded using a DL-based image codec, namely the JPEG AI standard [19], thus generating the color bitstream. The choice of JPEG AI as the image codec stems not only from its high compression efficiency, but also from being another learning-based JPEG standard, thus allowing to obtain a learning-based coding core for both geometry and color, effectively exploiting their synergies targeting a unified PC representation for man and machine.

On the receiver side, the JPEG PCC color decoding process works as follows:

- **2D Image Decoding**: The color bitstream is decoded using a DL-based image codec, namely the JPEG AI standard, which reconstructs the two projected 2D color images.
- **2D to 3D Inverse Projection**: Using the decoded geometry (from the geometry bitstream) and decoded images containing the projected color information, the inverse projection operation is performed to associate the decoded color information to the points in the decoded PC geometry.

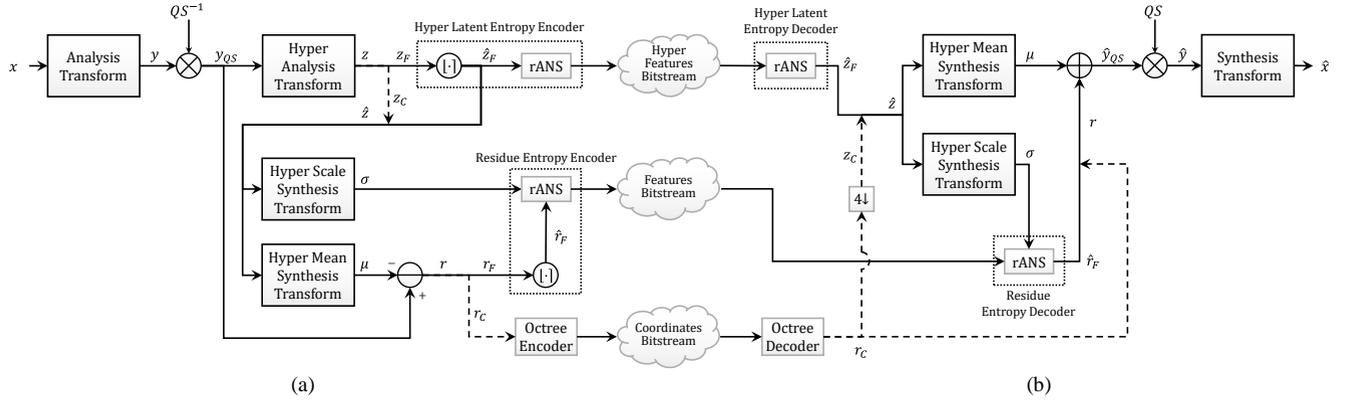

**FIGURE 3.** Overall architecture of the JPEG PCC DL coding model: (a) Encoder; and (b) Decoder. The dashed arrows represent only the coordinates of the sparse tensor.

- **Color Super-resolution**: Color coding is performed for the reconstructed PC points before performing up-sampling and SR. When geometry SR is performed (this is optional depending on the encoder decision), new geometry points are generated, which have no color. As such, a color SR module is used to assign a color to these newly generated points, producing the final decoded PC, now with geometry and color.

The following sections will detail the most important modules of the PC geometry and color encoding and decoding processes.

## B. DEEP LEARNING-BASED GEOMETRY CODING

This module is the heart of the JPEG PCC geometry coding pipeline, and its goal is to perform the actual encoding and decoding of the geometry data. It consists of an end-to-end DL coding model, which architecture was inspired by the successful works on learning-based image coding pioneered by Ballé *et al.* [9], namely using an autoencoder with a variational hyperprior.

The DL coding model processes a 3D PC geometry block using a sparse tensor representation [25]. Unlike a regular volumetric representation where the data is stored as a 3D matrix, in a sparse tensor representation, only the non-zero voxels are explicitly represented. This means that a data tensor $x$ is represented as a set of indices/coordinates of the non-zero voxels, $x_C$, and the corresponding voxel values/features, $x_F$, whereas the remaining elements are assumed to be zero. This is the type of representation for all the data passing through the DL coding model, from the input to the latent and hyper latent representations, as well as the residue; at the input stage, the non-zero voxel features take the value '1' thus signaling occupied voxels.

The DL coding model architecture is shown in Fig. 3, and Table 1 details the architecture of each of its modules, with each sequential layer/activation from the input to the output presented from top to bottom. SpConv corresponds to a sparse convolution layer and GTSpConv corresponds to a generative transposed sparse convolution layer, where the kernel size, number of input and output channels, and stride are shown between brackets, respectively. The encoder part of the DL coding model, shown in Fig. 3a, works as follows:

- **Analysis Transform**: The input block, $x$, is first transformed into a latent representation, $y$, by the analysis transform, i.e., the encoder part of the autoencoder model. Targeting a compact but rich representation, it consists of an alternating combination of sparse convolution layers and Inception-Residual Blocks (IRB) [31], with the former reducing the spatial dimensions of the input data and the latter extracting features from varying neighboring contexts with successively higher dimensionality. As detailed in Table 2, the IRB comprises multiple branches of sparse convolution layers with different kernel shapes and sizes, with the resulting output features being concatenated, and finally a residual skip connection from the IRB input to the concatenated features. The latent representation is then scaled by a user-defined QS (by default, QS = 1), resulting in $y_{QS}$.

- **Hyper Analysis Transform**: A hyper network is used to assist the entropy coding of the latent representation via forward adaptation. Unlike backward adaptation where the probability distributions are adapted based on the previously encoded symbols, in forward adaptation the probability distributions are adapted based on the actual symbols to encode, which requires including additional side information in the bitstream so the decoder can function. At the encoder side, the scaled latent representation, $y_{QS}$, is processed by the hyper analysis transform to capture its structural information, producing a hyper latent representation, $z$. The hyper analysis transform consists of three consecutive sparse convolution layers which extract features and reduce the spatial dimensions.

- **Hyper Latent Entropy Encoder**: This module encodes the hyper latent representation, which will serve as side information for entropy coding the latent representation, thus it also needs to be made available to the decoder.

**TABLE 1.** Detailed architecture of individual modules in the DL Coding Model.

| Analysis Transform | Synthesis Transform | Hyper Analysis Transform | Hyper Mean Synthesis Transform | Hyper Scale Synthesis Transform |
|---|---|---|---|---|
| SpConv($3^3$, 1, 32, 2↓) | GTSpConv($2^3$, 128, 128, 2↑) | SpConv($3^3$, 128, 128, 1) | GTSpConv($2^3$, 128, 128, 2↑) | GTSpConv($2^3$, 128, 128, 2↑) |
| ReLU | ReLU | ReLU | ReLU | ReLU |
| IRB(32) | IRB(128) | SpConv($3^3$, 128, 128, 2↓) | GTSpConv($2^3$, 128, 128, 2↑) | GTSpConv($2^3$, 128, 128, 2↑) |
| SpConv($3^3$, 32, 64, 2↓) | GTSpConv($2^3$, 128, 64, 2↑) | ReLU | ReLU | ReLU |
| ReLU | ReLU | SpConv($3^3$, 128, 128, 2↓) | SpConv($3^3$, 128, 128, 1) | SpConv($3^3$, 128, 128, 1) |
| IRB(64) | IRB(64) | | | |
| SpConv($3^3$, 64, 128, 2↓) | GTSpConv($2^3$, 64, 32, 2↑) | | | |
| ReLU | ReLU | | | |
| IRB(128) | IRB(32) | | | |
| SpConv($1^3$, 128, 128, 1) | SpConv($1^3$, 32, 1, 1) | | | |
| | Sigmoid | | | |
| **1208432 parameters** | **1127728 parameters** | **1327360 parameters** | **704896 parameters** | **704896 parameters** |

The hyper latent representation's features, $z_F$, are first quantized by simply rounding to integers, represented as $\lfloor \cdot \rceil$, resulting in $\hat{z}_F$. Then, $\hat{z}_F$ is entropy coded to generate the Hyper Features Bitstream. The used entropy coding engine is the range variant of Asymmetric Numeral Systems (rANS) [42] due to its high efficiency, combining a high compression ratio comparable to arithmetic coding with a low computational cost similar to Huffman coding. The $\hat{z}_F$ symbol's probability distributions are learned during the end-to-end training of the DL coding model.

- **Hyper Mean Synthesis Transform**: The quantized hyper latent representation, $\hat{z}$, (composed of coordinates, $z_C$, and features, $\hat{z}_F$) is processed by the hyper mean synthesis transform to produce a prediction, $\mu$, for the latent representation, $y_{QS}$. This prediction is then subtracted from the latent representation, $y_{QS}$, resulting in a residue, $r$. The hyper mean synthesis transform consists of three consecutive sparse convolution layers, in a symmetrical design to the hyper analysis transform.

- **Hyper Scale Synthesis Transform**: The quantized hyper latent representation, $\hat{z}$, is also processed by the hyper scale synthesis transform, which in turn estimates a set of standard deviations, $\sigma$, of the Gaussian distributions used to entropy code the residue, $r$. The hyper scale synthesis transform has the same architecture of the hyper mean synthesis transform.

- **Residue Entropy Encoder**: Finally, the residue, $r$, must be encoded. However, since the residue has a sparse tensor representation, both its features, $r_F$, and coordinates, $r_C$, have to be encoded. The features of the residue, $r_F$, are first quantized into $\hat{r}_F$ by simply rounding to integers, and then entropy coded to generate the Features Bitstream. The $\hat{r}_F$ symbol's probability distributions are assumed to be Gaussian with zero means and the standard deviations, $\sigma$, estimated by the hyper scale synthesis transform.

**TABLE 2.** Detailed architecture of the IRB and LIRB modules.

| | IRB (N channels) | LIRB (N channels) |
|---|---|---|
| **A** | SpConv(1×1×1, N/4, N/4, 1) | |
| | ReLU | |
| | SpConv(3×1×1, N/4, N/4, 1) | |
| | ReLU | |
| | SpConv(1×3×1, N/4, N/4, 1) | |
| | ReLU | |
| | SpConv(1×1×3, N/4, N/4, 1) | |
| | ReLU | |
| **B** | SpConv(1×1×1, N/4, N/4, 1) | SpConv(1×1×1, N/4, N/4, 1) |
| | ReLU | ReLU |
| | SpConv(3×3×3, N/4, N/4, 1) | SpConv(3×3×3, N/4, N/4, 1) |
| | ReLU | ReLU |
| | SpConv(1×1×1, N/4, N/4, 1) | SpConv(1×1×1, N/4, N/2, 1) |
| | ReLU | ReLU |
| **C** | SpConv(3×3×3, N/4, N/4, 1) | SpConv(3×3×3, N/4, N/4, 1) |
| | ReLU | ReLU |
| | SpConv(3×3×3, N/4, N/4, 1) | SpConv(3×3×3, N/4, N/2, 1) |
| | ReLU | ReLU |
| **D** | SpConv(5×5×5, N/4, N/4, 1) | |
| | ReLU | |
| | Concat(A, B, C, D) + Input | Concat(B, C) + Input |

- **Octree Encoder**: As for the coordinates, $r_C$, they are very critical and thus must be losslessly encoded to prevent impactful artifacts when reconstructing the sparse tensor. For this purpose, an octree encoder is used, namely the G-PCC standard [7] in lossless mode, generating the Coordinates Bitstream. It is worth noting that the hyper latent representation, $z$, is also a sparse tensor, thus the coordinates, $z_C$, should also be encoded; however, this is not necessary in practice since the coordinates, $z_C$, can be derived from the coordinates of the residue, $r_C$, at the decoder by simply down-sampling with a factor of four (equivalent to the two down-sampling sparse convolution layers of the hyper analysis transform).

The decoder part of the DL coding model, shown in Fig. 3b,

receives the three-part bitstream (Coordinates Bitstream, Features Bitstream, Hyper Features Bitstream) and generates the decoded PC block, $\hat{x}$, as follows:

- **Octree Decoder**: The Coordinates Bitstream is first decoded by the octree decoder, namely G-PCC Octree, producing the coordinates of the residue, $r_C$. These coordinates are then down-sampled by a factor of four to produce the coordinates of the hyper latent representation, $z_C$.

- **Hyper Latent Entropy Decoder**: The Hyper Features Bitstream is entropy decoded to produce the features of the hyper latent representation, $\hat{z}_F$. The $\hat{z}_F$ symbol's probability distributions are learned during the end-to-end training of the DL coding model. Together, $\hat{z}_F$ and $z_C$ form the quantized hyper latent representation $\hat{z}$.

- **Hyper Scale Synthesis Transform**: The quantized hyper latent representation, $\hat{z}$, is processed by the hyper scale synthesis transform to estimate the standard deviations, $\sigma$.

- **Residue Entropy Decoder**: The Features Bitstream is entropy decoded to produce the features of the residue, $\hat{r}_F$. The $\hat{r}_F$ symbol's probability distributions are assumed to be Gaussian with zero means and standard deviations, $\sigma$, as estimated by the hyper scale synthesis transform. Together, $\hat{r}_F$ and $r_C$ form the quantized residue $\hat{r}$.

- **Hyper Mean Synthesis Transform**: The quantized hyper latent representation, $\hat{z}$, is processed by the hyper mean synthesis transform to produce a prediction, $\mu$, which is then added to the quantized residue, $\hat{r}$, resulting in the decoded (quantized) latent representation, $\hat{y}_{QS}$.

- **Synthesis Transform**: The decoded latent representation is scaled back by the inverse QS, resulting into $\hat{y}$. Finally, the decoded latent representation, $\hat{y}$, is processed by the synthesis transform to produce the decoded block, $\hat{x}$, containing the probabilities of each voxel being occupied, with the final decoded coordinates being determined in the binarization module. The synthesis transform has a symmetrical architecture to the analysis transform.

The number of trainable parameters in each module of the DL coding model is shown in Table 1, with the total number of trainable parameters in the full DL coding model adding up to 5073312.

A common problem of DL-based codecs observed in the literature is the occurrence of catastrophic reconstruction errors due to mismatches in the probability distributions between encoder and decoder, even if minimal. This is especially prevalent when training, encoding, and decoding on different devices or machines, which may have different behaviors when dealing with floating-point representations, as is the case with most DL models. To mitigate this issue, all hyper scale synthesis transform operations were defined to be integer to guarantee bit-exact behavior. For this purpose, its layers' learnable parameters were quantized to integers by optimizing the parameters' integer precision while ensuring that the accumulator (i.e., the register storing the intermediate arithmetic operations' results) does not overflow [43]. To illustrate the importance of these quantized models, Fig. 4 shows an example of the type of decoding artifacts that can appear when encoding and decoding on different devices (GPU and CPU) and different machines. Without quantized models, it is clearly visible that some parts/blocks of the PC geometry are seemingly randomly reconstructed, completely losing the surface shape. This has a severe impact on the quality of the decoded PC, both subjectively and objectively, and is an unpredictable phenomenon as it may or may not occur for each block and its effect is different every time. As such, using quantized DL coding models is of utmost importance to guarantee an expected reconstruction, regardless of the device or machine.

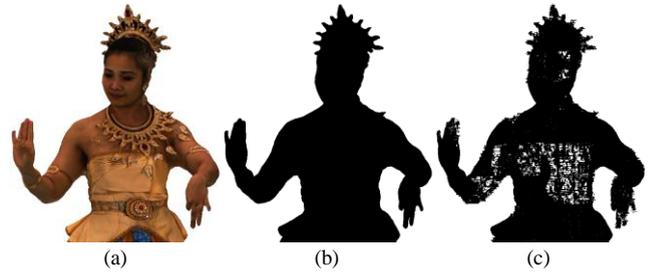

### C. DOWN-SAMPLING AND UP-SAMPLING
The goal of the down/up-sampling modules is to change the precision of the PC to facilitate efficient compression, especially in two specific conditions: first, it allows reaching lower rates, without severely penalizing the reconstruction quality; second, for sparser PC content, down-sampling the PC, making it denser, allows a better fit to the analysis transform, thus reaching lower encoding rates for the same reconstruction quality. This is because the voxel-based DL coding model may have difficulties when dealing with sparse PC content since the neighboring points are so far apart that the filter kernels in the convolutional layers cannot capture their dependencies.

At the encoder, down-sampling consists in simply down-scaling (i.e., dividing) the points' coordinates by the user-defined SF parameter, followed by rounding them to integers. Due to this, it is likely that multiple points end up in the exact same position, i.e., duplicate points, which are simply discarded (only one point is kept in each allowed position). The SF parameter is included in the bitstream.

At the decoder, after applying DL-based geometry decoding, up-sampling is performed to bring the PC back to its original precision, in the case down-sampling was performed at the encoder. This is achieved by simply up-scaling (i.e., multiplying) the points' coordinates by the SF included in the bitstream. This up-sampling operation does not change the number of decoded points as its goal is to simply change the precision.

### D. DEEP LEARNING-BASED GEOMETRY SUPER-RESOLUTION

The goal of the DL-based geometry SR module is to increase the density of the PC surface to improve the reconstruction quality. This module is important to achieve a better RD performance, especially for denser PCs, as performing SR at the decoder allows to recover most of the detail lost when down-sampling is applied at the encoder, thus significantly increasing the reconstruction quality with negligible rate cost (corresponding only to the $k_S$ parameter for the Top-k binarization of each block). The SR module can only be used after the up-sampling module if SF > 1; however, its usage is optional, being signaled via the $k_S$ parameter, namely if $k_S$ > 0. In practice, it is possible to apply up-sampling (after down-sampling), e.g., with SF = 2, and skip SR, if the encoder decides this is the best option making $k_S$ = 0.

The DL-based geometry SR module consists of a DL model with a U-net architecture [44] inspired by [45]. Like the DL coding model, it also processes a 3D PC block using a sparse tensor representation. As shown in the architecture in Fig. 5, the DL SR model consists of two paths:

- **Contracting Path**: Successively extracts features at different spatial resolutions, with five sets of one down-sampling sparse convolution layer followed by three Lightweight IRBs (LIRB), detailed in Table 2. The LIRB has a lower complexity than the IRB used in the DL coding model, consisting of only two branches instead of four, and with a lower number of channels in the intermediate layers. LIRBs are used instead of IRBs to keep the DL SR model's number of trainable parameters and computational complexity at an acceptable level.

- **Aggregating Path**: Aggregates the features from the different resolution levels, learning how to progress between resolutions in order to achieve the target resolution. At each resolution level, an up-sampling transposed sparse convolution is used to increase the resolution level, with the resulting features being concatenated with the corresponding features extracted in the contracting path, which then pass through three LIRBs and a sparse convolution layer. When up-sampling to the final resolution level, a generative transposed sparse convolution layer is used to generate the new points.

The difference between a regular transposed sparse convolution layer (TSpConv) and a generative transposed sparse convolution layer (GTSpConv) is that in the former its output points' coordinates are defined as the same as the input coordinates of the corresponding down-sampling SpConv layer in the contracting path, whereas in the latter new points are generated on all the kernel's positions while up-sampling. Two DL SR models are trained for different SF values, notably 2 and 4, since larger SF values were not shown to be useful for the PCs in the JPEG test dataset [46]. For SF = 2, the kernel size $m$ in the GTSpConv is set to 3, while for SF = 4 the kernel size $m$ is set to 5; this allows to generate more points when using a larger SF, thus providing a better reconstruction of the PC surface. The total number of trainable parameters in the full DL SR model for SF = 2 and SF = 4 add up to 7253817 and 7278905, respectively.

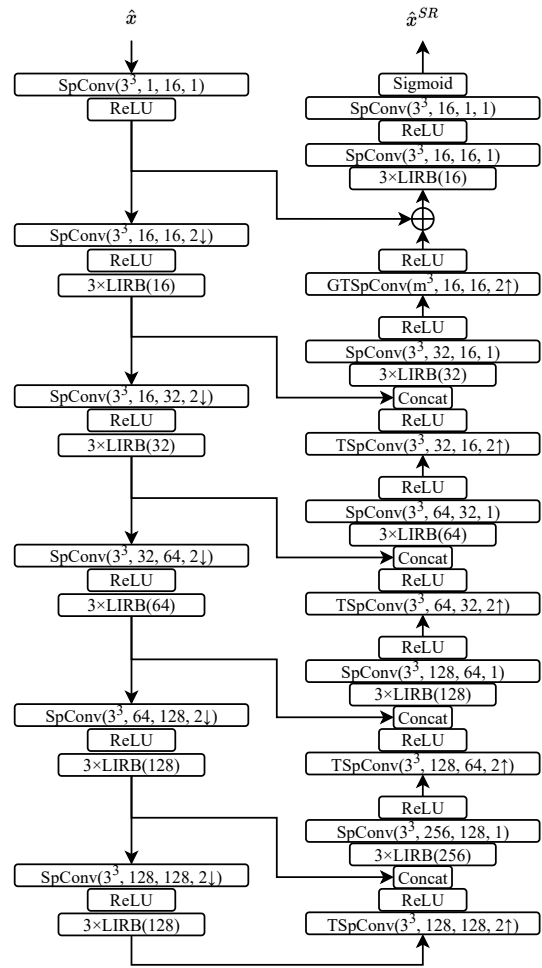

**FIGURE 5.** Architecture of the DL SR model.

### E. RECOLORING AND COLOR SUPER-RESOLUTION

The goal of these modules is to assign a color to the newly generated geometry points. The (non-normative) recoloring module is used when encoding the PC color, after encoding the PC geometry. Since the geometry codec is lossy, the number of points and their positions in the decoded PC may change regarding the original PC. However, since the PC color cannot exist without the underlying PC geometry, it is necessary to attribute a color to the new points produced by the decoder; this requires a recoloring approach to transfer the color from the original PC to the decoded PC geometry. This recoloring does not take into account the DL-based geometry SR module as it generates many new points, which color information would have to be encoded, leading to a significant increase in rate. Instead, the points generated by the DL-based geometry SR module are assigned a color at the decoder via the (normative) color super-resolution module, with no rate cost, considering only the colors of the neighboring points in the decoded PC.

For both these modules, the same approach is applied as follows:

- For collocated points, i.e., points with the same coordinates in both reference and target PCs, a simple copy of the color from the reference PC to the target PC is performed.
- For points in the target PC without a direct correspondence in the reference PC, their color is determined using a Radial Basis Function interpolation [47] with a linear kernel considering the 20 nearest neighbors in the reference PC; naturally, the result of the color interpolation is rounded to 8-bit unsigned integers.

### F. 3D TO 2D PROJECTION AND 2D TO 3D INVERSE PROJECTION

Given the recolored PC, the goal of these modules is to project the 3D PC color information onto 2D images and vice-versa. For this purpose, the projection approach used by the MPEG V-PCC standard [7] was adopted in JPEG PCC, due to its proven efficiency. The 3D to 2D projection module can be described by the following steps [7]:

- **Patch Generation**: The PC is decomposed into a set of independent 3D patches as follows:
  - **Initial Segmentation**: Given the normal vector for each point, an initial clustering is obtained by associating each point to the bounding box plane which has the closest normal vector.
  - **Refined Segmentation**: The initial clustering is then iteratively refined based on the neighborhood of each point with the goal to generate smoother clusters.
  - **Patch Segmentation**: Patches are extracted from the clusters based on a connected components method, creating contiguous patches.
- **Patch Projection and Packing**: The 3D patches are mapped to 2D using an orthogonal projection. Then, from the largest to the smallest, the 2D patches are placed onto a 2D image on the first position that guarantees no overlapping, following a raster scan order; if a position cannot be found, then the 2D image height is doubled. Finally, the 2D image height is trimmed after packing all patches to remove the empty space.
- **Image Generation**: An image containing the color information is generated using the RGB values of the 3D points in the corresponding RGB channels of the image; naturally, the geometry image and occupancy map usually produced by V-PCC are ignored in JPEG

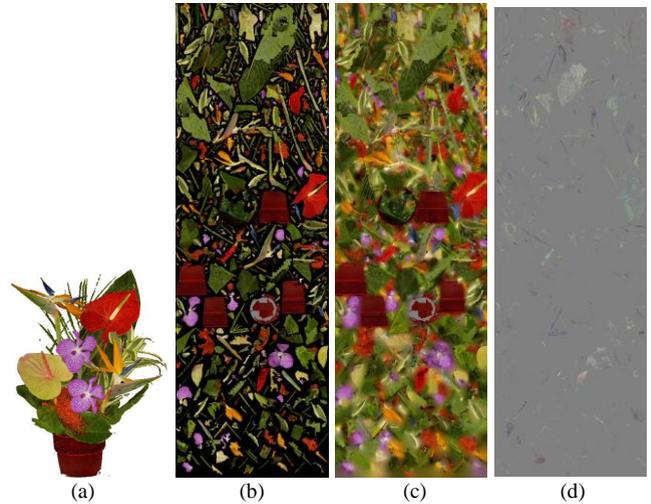

**FIGURE 6. Example of images resulting from 3D to 2D Projection: (a) Original *Bouquet* PC; (b) Generated near layer image, before filling the background pixels; (c) Generated near layer image, after applying the smoothed push-pull algorithm to fill the background pixels; (d) Generated differential far layer image, after trimming.**

PCC since the geometry information has already been decoded at this stage using the DL-based geometry coding module. Since a patch may have points projected onto the same 2D pixel, two layers are used to store the overlapping points, where the first (near) layer contains the points with lowest depth and the second (far) layer contains the points with highest depth. Then, a smoothed push-pull padding algorithm [6] is used to fill the background pixels of the images to generate smoother images for a more efficient coding. Since the images corresponding to the two layers tend to be rather similar, the near layer keeps the absolute color values while the far layer becomes differential with respect to the near layer by computing the residue, followed by 8-bit quantization. Finally, the differential far layer image is trimmed by removing the outer edges with zero-valued residue pixels, thus removing pixels with no information to further reduce the required rate. The position of the trimmed image is included in the bitstream.

Fig. 6 shows an example of generated images, before (Fig. 6b) and after (Fig. 6c) the background filling process, where it is noticeable that the background filling produces a considerably smoother image; Fig. 6d shows the corresponding differential far layer image, demonstrating that it contains significantly less information than the absolute near layer image.

At the decoder, the 2D to 3D inverse projection module is used to reconstruct the PC with color, and can be described by the following steps:

- **3D to 2D Geometry Projection**: Since the geometry information is decoded first in a separate pipeline, the steps described in the 3D to 2D projection module are performed again on the decoded PC geometry as at the encoder, producing the projected geometry images, occupancy map, and patch information (i.e., the location of each patch in the 2D images and the 3D space, and the corresponding bounding box plane used for projecting), all required to be able to reconstruct the PC from the projected images.

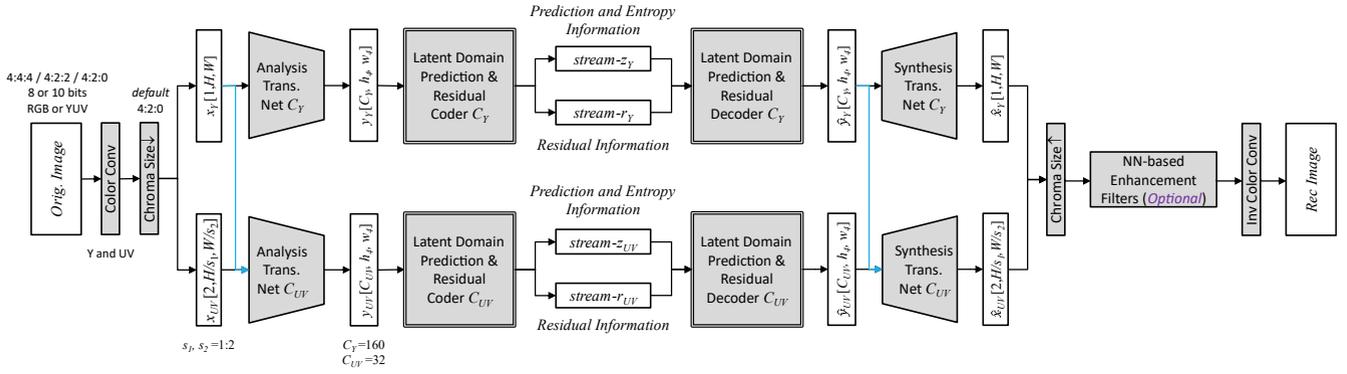

**FIGURE 7.** Overall architecture of the JPEG learning-based image coding standard. The end-to-end codec processes the luminance (Y) and chrominance (UV) components separately in two parallel pipelines [48].

- **PC Reconstruction**: Each patch is placed in the corresponding bounding box plane and position, as defined in the patch information, and the binary occupancy map is used to determine which pixels in the corresponding projected images actually correspond to points in the PC. For each occupied pixel, the pixel's coordinates in the corresponding 2D image will be assigned as the two 3D point's coordinates in the same plane, whereas the pixel's luminance value will be assigned as the third coordinate (i.e., in the orthogonal axis), reconstructing the PC geometry. Finally, the corresponding pixel's RGB values in the images containing the color are directly assigned to the point, thus reconstructing the PC color.

### G. 2D IMAGE CODING
The goal of this module is to code the projected and packed PC color images (absolute near layer and differential far layer), using the JPEG learning-based image coding standard, designated JPEG AI [19]. The overall JPEG AI high-level coding architecture [48] is presented in Fig. 7. This architecture has evolved from several earlier works on learning-based image coding in the literature and built upon an autoencoder with a variational hyperprior architecture, thus bearing a resemblance to the DL coding model for geometry coding described in Section III.B. Since JPEG AI is not the focus of this paper, only a high-level description of its DL coding model is provided here; more details can be found in [48]. The main characteristics of the JPEG AI codec include:

- **Component Separation**: First, the RGB image is converted into the YUV format, where the chrominance components may be down-sampled to take advantage of the Human Visual System's characteristics to achieve higher compression efficiency, as performed in most conventional image codecs in the past. Then, the color components are coded separately, using one neural network to code the luminance (Y) and another to code the chrominances (UV); both neural networks have identical architectures, differing on the number of filters, with the latent representations containing 160 and 32 channels for the luminance and chrominances, respectively; the network for the luminance component is deeper due to the higher spatial resolution. Furthermore, the chrominances are conditionally coded considering the previously coded luminance information.

- **Analysis and Synthesis Transforms**: The analysis and synthesis transforms form an autoencoder architecture. To offer flexibility in terms of the RD performance and computational complexity trade-off, JPEG AI defines two (non-normative) analysis and three (normative) synthesis transforms, all interoperable with each other [48], with the most complex ones including a transformer-based attention module and a convolution-based attention block, which allow reaching higher compression efficiency. Early 2025, JPEG AI defined three hierarchical profiles, named Simple, Base and High, with increasing complexity and compression efficiency.

- **Latent Domain Prediction and Residual Coder**: Similar to the DL coding model described in Section III.B, a hyper encoder and a hyper decoder are used to create a prediction for the latent representation; a hyper scale decoder is used to estimate the probability distribution parameters to entropy code the residue. Moreover, JPEG AI additionally includes a multi-stage context modelling network and a prediction fusion network to improve the latent prediction produced by the hyper decoder.

- **Variable Rate Coding**: An extrapolation gain unit is used to allow variable rate coding, where a gain vector is determined based on the models trained for different RD trade-offs; this allows to extrapolate different RD trade-offs from those learned during training. Each element of the gain vector corresponds to a scale factor for each of the channels in the latent residue; at inference time, the latent residue is multiplied by the gain vector.

- **Optional Tools**: JPEG AI additionally includes several optional tools for improving the compression efficiency and reconstruction quality by providing content adaptation. These include post-processing enhancement filters, amplification of selected latent residue channels, and scaling certain elements of the latent representation; these tools are all optimized at the encoder with auxiliary information sent to the decoder.

In JPEG PCC, JPEG AI is used to encode both projected images. The two bitstreams produced at the encoder (one for each projected image) are packed together to form the color bitstream, as described earlier.

### H. GEOMETRY PIPELINE TRAINING PROCESS

This section describes the process and conditions used for training the DL coding and the DL SR models used in the geometry coding pipeline. It is important to observe that the training process and conditions are not normative, but the decoder model parameters are normative and thus specified in a table included in the standard. Though they have many things in common, the models for coding and SR serve different purposes, and therefore some training conditions are slightly different, as detailed in the following.

#### 1) DATASET

The dataset is composed by a selection of static PCs listed in the JPEG Pleno PCC Common Training and Testing Conditions (CTTC) [46], split into training and validation sets, as shown in Table 3. Some of the sparser PCs were down-sampled from their original precision to a lower one (as indicated in Table 3 by the $\rightarrow$ symbol). The PCs were partitioned into blocks. The blocks with a very low point count (i.e., less than 500 points) were discarded to avoid having a negative impact on the training process. For the coding models and the SR model for SF = 2, blocks of size 64×64×64 were used, totaling 35303 blocks for training and 3421 blocks for validation, whereas for the SR model for SF = 4, blocks of size 128×128×128 were used, totaling 9572 blocks for training and 993 blocks for validation.

#### 2) LOSS FUNCTION: CODING MODEL

Since the goal of JPEG PCC stage 1 is human visualization, compression efficiency was the main goal, thus the natural loss function to train the DL coding model follows a traditional RD formulation with a Lagrangian multiplier, $\lambda$:

$$Loss\ Function = Distortion + \lambda \times Estimated\ Rate. \quad (1)$$

During training, there is no actual quantization and entropy coding performed as it would not be differentiable, which is a critical requirement to perform the training algorithm; thus, the coding rate of each block is estimated considering the entropy of the latent residue and the hyper latent representation, according to their respective probability distributions. As for the distortion term, an error between the original input block and the decoded block should be measured. However, it is not possible to use common distance-based PC distortion metrics from the literature since the output of the models is a set of probabilities of each voxel being occupied, thus requiring a binarization process to determine the decoded PC's coordinates, which is non-differentiable. As such, the geometry distortion is measured as a binary classification error for each voxel, averaged across the entire block, using the Focal Loss [49] defined as:

$$FL(v,u) = \begin{cases} -\alpha(1-v)^\gamma \log v, & u = 1 \\ -(1-\alpha)v^\gamma \log(1-v), & u = 0 \end{cases}, \quad (2)$$

where $u$ is the original voxel binary value and $v$ is the corresponding decoded voxel probability. A weight $\alpha$ is used to mitigate the imbalance between the number of '0' and '1' valued voxels. The weight $\gamma$ is used to give more focus to still misclassified voxels as opposed to those already correct. The values $\alpha = 0.5$ and $\gamma = 2$ were found to be appropriate during the development of the standard. A total of five coding models were trained to reach different RD trade-offs, using $\lambda = 0.05$, 0.025, 0.01, 0.005, and 0.0025; training was performed sequentially, from lowest to highest $\lambda$ [29], i.e., using the previous model as initialization.

**TABLE 3.** Training and validation datasets for the DL Coding and SR Models.

| | Point Cloud | Precision (bits) | Number of Points | Number of $64^3$ Blocks | Number of $128^3$ Blocks |
|---|---|---|---|---|---|
| **Training** | *Loot* | 10 | 805285 | 192 | 49 |
| | *Redandblack* | 10 | 757691 | 166 | 46 |
| | *Longdress* | 10 | 857966 | 178 | 44 |
| | *Andrew10* | 10 | 1276312 | 224 | 68 |
| | *David10* | 10 | 1492780 | 277 | 81 |
| | *Phil10* | 10 | 1660959 | 295 | 87 |
| | *Ricardo10* | 10 | 960703 | 170 | 47 |
| | *Sarah10* | 10 | 1355867 | 258 | 75 |
| | *The20sMaria* | 11 | 3681165 | 669 | 175 |
| | *UlliWegner* | 10 | 537042 | 150 | 39 |
| | *BasketballPlayer* | 11 | 2925514 | 682 | 184 |
| | *Exercise* | 11 | 2391718 | 530 | 137 |
| | *Model* | 11 | 2458429 | 561 | 129 |
| | *Mitch* | 11 | 2289640 | 821 | 218 |
| | *ThomasSenic* | 11 | 2277443 | 749 | 188 |
| | *Football* | 11 | 1021107 | 160 | 42 |
| | *Facade15* | 14 → 12 | 6834258 | 2517 | 761 |
| | *Facade64* | 14 → 12 | 12755151 | 3539 | 965 |
| | *EgyptianMask* | 12 → 9 | 269739 | 141 | 43 |
| | *Head00039* | 12 | 13903516 | 9218 | 3044 |
| | *Landscape00014* | 14 → 12 | 15270319 | 3069 | 834 |
| | *StanfordArea2* | 16 → 12 | 19848824 | 4178 | 882 |
| | *StanfordArea4* | 16 → 12 | 24236048 | 6559 | 1434 |
| **Validation** | *Queen* | 10 | 1000993 | 179 | 49 |
| | *Dancer* | 11 | 2592758 | 604 | 161 |
| | *Frog00067* | 12 → 11 | 3321097 | 1467 | 454 |
| | *IpanemaCut* | 12 → 11 | 4938418 | 1171 | 329 |

### 3) LOSS FUNCTION: SR MODEL

For the SR model, since it has no effect on the coding rate, the loss function consists only of the distortion term, which uses the same Focal Loss metric. However, unlike the coding models, for the SR model the ground truth is not the input block but rather the original block before being down-sampled. Furthermore, the SR models were trained without considering coding, thus each SR model is used for all RD trade-offs, requiring only a total of two SR models, trained for SF = 2 and SF = 4. While experiments have been performed where a SR model was trained for each RD trade-off ($\lambda$) coding model, considering as input the decoded PC block, no significant advantage was found in terms of RD performance; for this reason, only two SR models (for SF = 2 and SF = 4) are used, independently of the target rate.

### 4) HYPER-PARAMETERS

All geometry coding and SR models were implemented and trained in PyTorch version 1.13, using the Minkowski Engine v0.5.4 [50] for the sparse tensor representation and sparse convolutions. The Adam algorithm [51] was used for optimization, with an initial learning rate set to $10^{-4}$, reduced to $10^{-5}$ when the validation loss does not decrease for 10 consecutive epochs, within a 0.1% tolerance margin. To prevent overfitting, an early stopping criterion was used with a patience of 25 epochs and a tolerance margin of 0.1%. Minibatches of 8 and 4 blocks were used for the coding and SR models with SF = 2 and for the SR model with SF = 4, respectively.

### I. COLOR PIPELINE TRAINING PROCESS

Regarding the color coding pipeline, JPEG AI is the only trainable module. JPEG AI is used directly in JPEG PCC, meaning its training process and conditions were not changed or adapted. Since JPEG AI is not the focus of this paper, the training process and conditions are only briefly described here.

### 1) DATASET

The dataset used to train JPEG AI consists of over 5000 natural images and JPEG PCC uses the JPEG AI standard models trained with these images [52]. However, the JPEG PCC projected images to be coded, both absolute and differential, as shown in Fig. 6, are very different from natural images, what may create some inefficiency.

### 2) LOSS FUNCTION

Similar to the DL coding model for geometry coding, the JPEG AI training loss function also follows a RD formulation. The training process follows multiple stages, using different distortion metrics in the loss function. The first stages use the Mean Squared Error (MSE) as the distortion metric, whereas latter stages combine MSE and Multi-Scale Structural Similarity Index Measure (MS-SSIM) distortions [53].

## IV. PERFORMANCE ASSESSMENT

In this section, the performance of the JPEG Pleno Learning-based PC Coding standard is assessed in comparison with other PC coding standards and relevant DL-based solutions in the literature, all already briefly reviewed in Section II.

### A. EXPERIMENTAL SETUP

In order to properly assess JPEG PCC's compression performance, it is necessary to establish an appropriate experimental setup, namely considering a representative test dataset, meaningful performance metrics, and relevant benchmarks.

### 1) TEST DATASET

The test dataset, shown in Fig. 8 and detailed in Table 4, is composed by 12 static PCs listed in the JPEG Pleno PCC CTTC [46], selected with the goal of using a diverse set of objects/scenes with different characteristics, notably in terms of precision (bit depth), density, homogeneity, and color gamut volume. Following the same definitions as MPEG [54], PCs are grouped into three categories according to their density (measured by the density factor), namely solid, dense, and sparse, since this is one of the PC characteristics with the biggest impact in compression efficiency. The density and homogeneity factors are determined by first computing the local density around each point in the PC, given by the number of points inside the smallest sphere containing at least its 12 nearest neighbors divided by the sphere's volume. Then the density factor is computed as the negative logarithm of the median of all local densities. A PC with a density factor lower than 1 is considered "solid", greater than 2 is considered "sparse", and in between is considered "dense" [54].

Regarding homogeneity, each PC is considered homogeneous or heterogeneous depending on whether the local densities are consistent across all the PC or whether they vary more significantly, respectively. The homogeneity factor is defined as the difference between the third and first quartiles of the local densities, normalized by the local densities range. A PC with a homogeneity factor lower than 11.6 is considered "homogeneous"; otherwise, it is considered "heterogeneous" [54].

Regarding the color, the richness of the color information for each PC is measured through the color gamut volume, defined as the percentage of the color space volume occupied by the convex hull of the distribution of all points' colors, for which a higher value means that a higher number of different colors is present in the PC.

### 2) PERFORMANCE METRICS

To assess the RD performance, several metrics have been considered, following the JPEG Pleno PCC CTTC [46], namely:

- **Geometry rate**: Measures the rate for the PC geometry only, computed as the number of bits in the geometry bitstream divided by the total number of points in the original PC, represented as bits-per-point (bpp).
- **Total rate**: Measures the total rate for the PC geometry and color, computed as the number of bits in the geometry and color bitstreams divided by the total number of points in the original PC, represented as bpp.

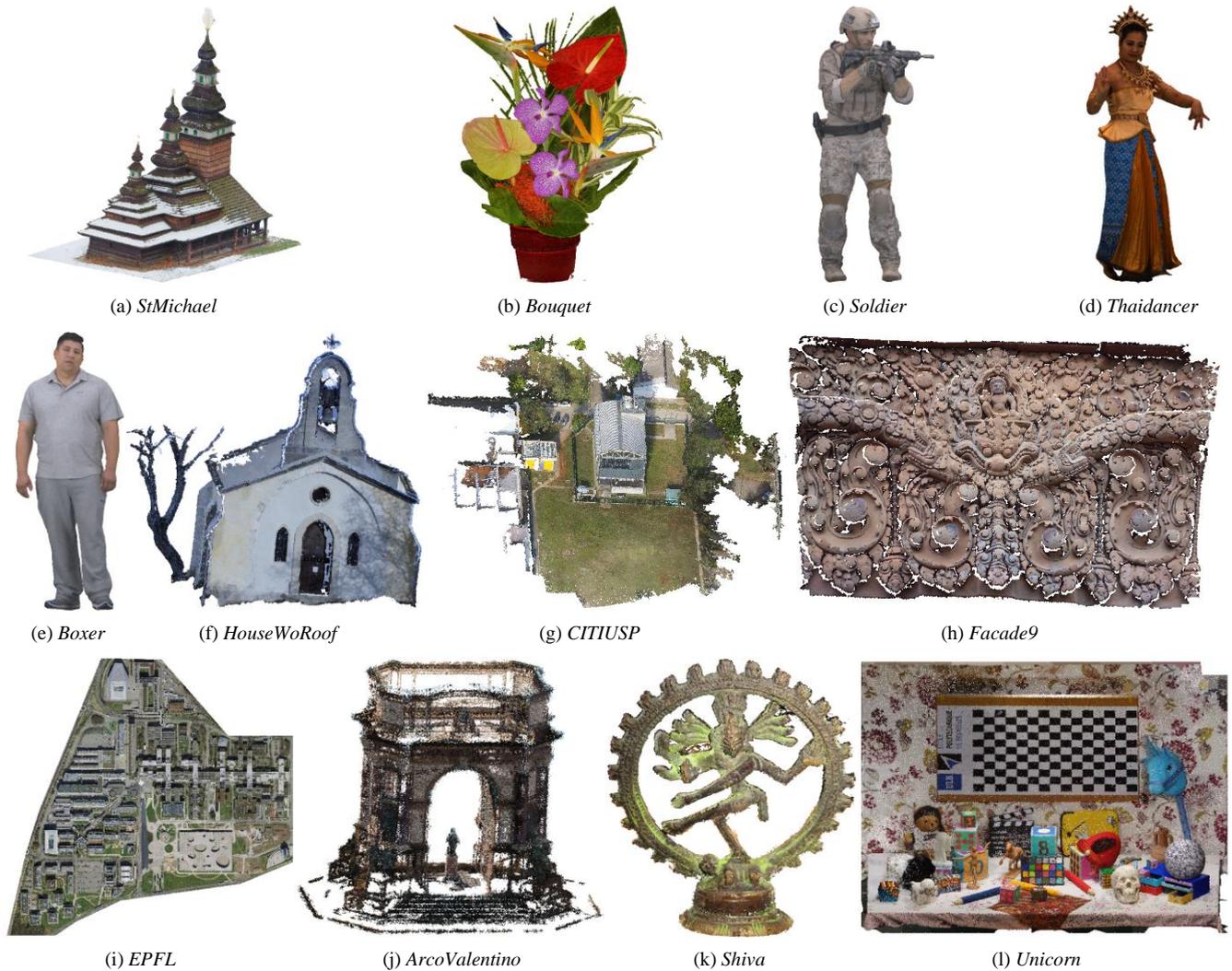

| (a) *StMichael* | (b) *Bouquet* | (c) *Soldier* | (d) *Thaidancer* |

| (e) *Boxer* | (f) *HouseWoRoof* | (g) *CITIUSP* | (h) *Facade9* |

| (i) *EPFL* | (j) *ArcoValentino* | (k) *Shiva* | (l) *Unicorn* |

**FIGURE 8.** PC test dataset to evaluate the RD performance of the JPEG Pleno PCC standard, as defined in the JPEG CTTC [46].

**TABLE 4.** Test dataset for assessing the JPEG PCC performance.

| | Point Cloud | Bit Depth | No. Points | Density Factor | Homogeneity Factor | Color Gamut Volume |
|---|---|---|---|---|---|---|
| **Solid** | *StMichael* | 10 | 1871158 | 0.379 | 6.897 (Hom) | 21% |
| | *Bouquet* | 10 | 3150249 | 0.379 | 10.345 (Hom) | 41% |
| | *Soldier* | 10 | 1089091 | 0.379 | 8.333 (Hom) | 1% |
| | *Thaidancer* | 12 | 3130215 | 0.484 | 5.000 (Hom) | 22% |
| **Dense** | *Boxer* | 12 | 3493085 | 1.314 | 11.111 (Hom) | 3% |
| | *HouseWoRoof* | 12 | 4848745 | 1.438 | 37.415 (Het) | 13% |
| | *CITIUSP* | 13 | 5705126 | 1.576 | 22.642 (Het) | 29% |
| | *Facade9* | 12 | 1596085 | 1.856 | 13.208 (Het) | 5% |
| **Sparse** | *EPFL* | 13 | 4694733 | 2.002 | 17.390 (Het) | 33% |
| | *ArcoValentino* | 12 | 1481746 | 2.385 | 7.066 (Hom) | 15% |
| | *Shiva* | 12 | 1009132 | 2.424 | 11.304 (Hom) | 19% |
| | *Unicorn* | 13 | 1995189 | 3.282 | 12.500 (Het) | 51% |

- **PSNR D1**: Measures the geometry quality of the decoded PC by considering the point-to-point distance between the decoded and original PCs in both directions (i.e., from decoded to original and from original to decoded); the point-to-point distance is computed as the average squared distance between each point in the target PC and its nearest corresponding point in the reference PC, taking the maximum between the two directions.

- **PSNR D2**: Measures the geometry quality of the decoded PC by considering the point-to-plane distance between the decoded and original PCs in both directions; the point-to-plane distance is computed similarly to the point-to-point distance, but it considers the projection of each distance onto the normal vector of the corresponding original PC point [55].

- **PSNR Y**: Measures the color quality of the decoded PC by considering the MSE of the luminance between the decoded and original PCs in both directions; the MSE is computed as the average squared error between the luminance of each point in the target PC and its nearest corresponding point in the reference PC, taking the maximum between the two directions.

- **PSNR YUV**: Measures the color quality of the decoded PC by considering the MSE of the luminance and chrominances between the decoded and original PCs in both directions; the PSNR YUV is computed as a weighted average of the PSNR Y, PSNR U, and PSNR V (computed the same way as PSNR Y, but with the corresponding chrominance), with the weight of luminance being six times that of each chrominance.
- **1-PCQM**: Measures the joint geometry and color quality of the decoded PC by considering a linear combination of several geometry-based and color-based features computed between the decoded and original PCs [56].

In addition to the described metrics, the Bjontegaard-Delta Rate (BD-Rate) and Bjontegaard-Delta PSNR (BD-PSNR) are used to quantify the gains/losses between two codecs, with negative BD-Rate and positive BD-PSNR implying better RD performance than the reference codec.

### 3) BENCHMARKS
The relevant benchmarks to assess the JPEG PCC RD performance include the MPEG PC coding standards, as well as state-of-the-art DL-based PC geometry coding solutions with publicly available software. The benchmarks and their specific coding configurations are detailed as follows:

- **G-PCC Octree v24 / G-PCC Octree PredLift v24**: MPEG G-PCC standard reference software version 24 (available in: https://github.com/MPEGGroup/mpeg-pcc-tmc13) was used with the Octree geometry coding mode and the PredLift color coding mode, with the specific coding configurations for each PC as specified in the JPEG Pleno PCC CTTC [46].
- **V-PCC Intra v23**: MPEG V-PCC standard reference software version 23 (available in: https://github.com/MPEGGroup/mpeg-pcc-tmc2) was used with VVC [23] in the Intra mode, with the specific coding configurations for each PC as specified in the JPEG Pleno PCC CTTC [46].
- **PCGCv2 [25]**: PCGCv2 codec (available in: https://github.com/NJUVISION/PCGCv2) was used with the default coding configurations as specified in the software, namely considering the parameters "--scaling_factor=1.0 --rho=1.0" for solid PCs, "--scaling_factor=0.375 --rho=1.0" for dense PCs, and "--scaling_factor=0.375 --rho=4.0" for sparse PCs.
- **PCGFormer [26]**: PCGFormer codec (available in: https://github.com/3dpcc/PCGFormer) was used with the default coding configurations as specified in the software, namely considering the same parameters as specified for PCGCv2.
- **GRASP-Net [27]**: GRASP-Net codec (available in: https://github.com/InterDigitalInc/GRASP-Net) was used with the default coding configurations as specified in the software, namely considering for each PC the specific models trained for the corresponding PC category.

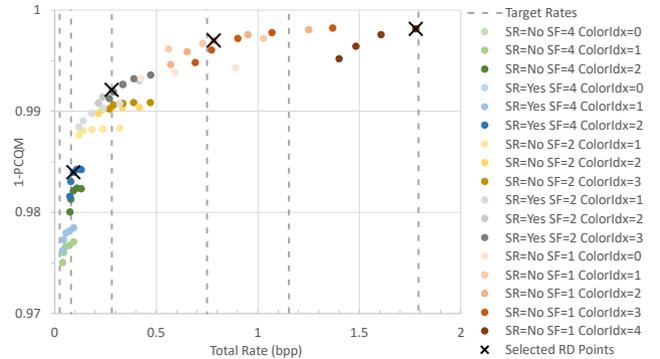

**FIGURE 9. Example of the best JPEG PCC coding configurations selection for the *Soldier* PC. Each dot represents one JPEG PCC coding configuration; dots with the same color only vary the *GeoIdx* parameter. From the six target rates (represented by the vertical dashed lines), four RD points were selected such that the 1-PCQM metric was maximized (marked with an "X").**

Although SparsePCGC [32] is a direct upgrade over PCGCv2 [25], there is no publicly available software, and thus it could not be used as a benchmark. While the MPEG PCC standards allow both geometry-only as well as geometry and color coding, the DL-based solutions listed above offer solely geometry-only coding. Despite the significant progress and performance achieved in the DL-based PC geometry coding field, DL-based PC color coding is still fairly underdeveloped in the literature, with no publicly available solutions.

### 4) JPEG PCC CODING CONFIGURATIONS
In the following experiments, the JPEG PCC VM software version 4.1 (April 2024) was used. As described in the previous section, JPEG PCC has several coding parameters that can impact the RD performance, namely the block size (BS), the sampling factor (SF), the quantization step (QS), the usage of SR, the geometry rate index (GeoIdx), and the color rate index (ColorIdx). The GeoIdx takes values between 0 and 4, signaling which DL coding model to use for geometry coding, where 0 corresponds to the model trained for lowest rate (largest $\lambda$) and 4 corresponds to the model trained for highest rate (smallest $\lambda$). The ColorIdx is equivalent to GeoIdx to select the JPEG AI coding model used for color coding.

To provide a fair comparison with the benchmarks, a set of target rates was defined considering the MPEG G-PCC benchmark as a baseline, as specified in the JPEG Pleno PCC CTTC [46]. The JPEG PCC configurations were selected for each PC in the test dataset by searching for the best coding configurations for each PC, testing the different combinations of coding parameters, and selecting the one providing the best joint quality (considering the 1-PCQM joint metric) for each target rate; only four target rates for each PC were considered, since for some PCs it was not possible to reach all target rates [46].

Fig. 9 shows an example of the selection process for the best JPEG PCC coding configurations, maximizing the 1-PCQM

**TABLE 5.** JPEG PCC coding configurations for each PC in the test dataset.

| | Point Cloud | R1 | | | | R2 | | | | R3 | | | | R4 | | | |
|---|---|---|---|---|---|---|---|---|---|---|---|---|---|---|---|---|---|
| | | SF | SR | GeoIdx | ColorIdx | SF | SR | GeoIdx | ColorIdx | SF | SR | GeoIdx | ColorIdx | SF | SR | GeoIdx | ColorIdx |
| **Solid** | *StMichael* | 4 | Yes | 1 | 1 | 2 | Yes | 2 | 2 | 1 | No | 2 | 2 | 1 | No | 4 | 4 |
| | *Bouquet* | 4 | Yes | 4 | 1 | 1 | No | 2 | 0 | 1 | No | 3 | 2 | 1 | No | 4 | 4 |
| | *Soldier* | 4 | Yes | 2 | 2 | 2 | Yes | 2 | 2 | 1 | No | 2 | 2 | 1 | No | 3 | 4 |
| | *Thaidancer* | 4 | Yes | 1 | 0 | 2 | Yes | 1 | 1 | 1 | No | 1 | 2 | 1 | No | 3 | 4 |
| **Dense** | *Boxer* | 4 | Yes | 0 | 1 | 2 | Yes | 1 | 1 | 2 | Yes | 4 | 3 | 1 | No | 3 | 3 |
| | *HouseWoRoof* | 4 | Yes | 1 | 1 | 2 | Yes | 2 | 2 | 1 | No | 3 | 2 | 1 | No | 4 | 3 |
| | *CITIUSP* | 4 | No | 0 | 0 | 4 | No | 3 | 2 | 2 | No | 3 | 1 | 2 | Yes | 4 | 3 |
| | *Facade9* | 4 | Yes | 1 | 2 | 2 | Yes | 2 | 3 | 1 | No | 3 | 2 | 1 | No | 4 | 2 |
| **Sparse** | *EPFL* | 4 | No | 0 | 0 | 4 | No | 3 | 1 | 2 | No | 3 | 1 | 2 | No | 4 | 3 |
| | *ArcoValentino* | 4 | No | 0 | 0 | 4 | No | 2 | 1 | 2 | No | 3 | 1 | 2 | No | 4 | 3 |
| | *Shiva* | 4 | No | 1 | 1 | 2 | No | 2 | 2 | 2 | Yes | 4 | 3 | 1 | No | 3 | 2 |
| | *Unicorn* | 4 | No | 1 | 0 | 4 | No | 3 | 2 | 4 | No | 4 | 2 | 2 | No | 3 | 2 |

quality metric. In terms of the geometry coding parameters, the BS and QS were fixed to 128 and 1, respectively, since they have a less critical impact on the RD performance, while the SF, the usage of SR, and GeoIdx were optimized. In terms of color coding, the JPEG AI VM software version 5.1 was used with the High profile, and the Tools On configuration. The ColorIdx parameter was optimized for each PC and target rate. Table 5 shows the JPEG PCC coding configurations resulting from the optimization for each PC and rate point, determined using the previously explained approach. These configurations will be used in the following experiments.

### B. GEOMETRY-ONLY RD PERFORMANCE

This sub-section assesses the JPEG PCC RD performance when only the PC geometry is coded. Following the previously described experimental setup, the obtained results are presented in Fig. 10, showing RD charts with the average performance of the codecs for each PC category (solid, dense, and sparse), for both PSNR D1 and PSNR D2 metrics. Moreover, Table 6 details the BD-Rate and BD-PSNR results, comparing JPEG PCC with each of the considered benchmarks, used as reference; results in bold represent the cases when JPEG PCC outperforms the benchmark. These results allow the following observations:

**Solid PCs**

- Compared to the conventional MPEG standards, JPEG PCC achieves a considerably better RD performance for all solid PCs, reaching an average BD-Rate of -87% (-80%) and -49% (-47%) over G-PCC Octree and V-PCC Intra, respectively, for PSNR D1 (PSNR D2). The G-PCC Octree performance for this type of PCs tends to be significantly poorer than the remaining benchmarks, since increasing the G-PCC Octree quantization step causes a reduction on the number of points, which degrades the detail and density of the PC's surface.

- Compared to the DL-based benchmarks, JPEG PCC outperforms all three codecs on average, with a BD-Rate of -5% (10%), -17% (-8%), and -30% (-34%) over PCGFormer, PCGCv2, and GRASP-Net, respectively, for PSNR D1 (PSNR D2); the exception seems to be PCGFormer for PSNR D2, which presents a slightly better RD performance. While GRASP-Net falls behind the other two codecs, on average, the advantage of PCGCv2 and PCGFormer over GRASP-Net and even JPEG PCC seems to be limited to the 10-bit precision PCs, with the *Bouquet* PC showing particularly higher gains (mainly at lower rates).

**Dense PCs**

- Compared to the conventional MPEG standards, JPEG PCC achieves better average RD performance than G-PCC Octree for all dense PCs, and V-PCC Intra for the *Boxer* and *HouseWoRoof* PCs. For the *Facade9* PC, V-PCC Intra has a massive BD-Rate advantage over JPEG PCC, as well as the remaining codecs. This result, which dominates the average for dense PCs, can be justified by the particular PC characteristics (a single façade with detailed relief) which perfectly suit the projection-based codec, allowing it to be extremely efficient. For this type of content, while G-PCC Octree still underperforms the remaining codecs at low and medium rates, at high rates it is advantageous since it is capable of achieving increasing quality, unlike the other codecs that tend to saturate the reconstruction quality.

- Compared to the DL-based benchmarks, JPEG PCC consistently outperforms both PCGCv2 and PCGFormer for all dense PCs, with an average BD-Rate of -60% (-38%) and -72% (-22%), respectively, for PSNR D1 (PSNR D2). On the other hand, compared to GRASP-Net, JPEG PCC achieves only a -4% (-21%) average BD-Rate, with a better RD performance for *Boxer* and *HouseWoRoof* PCs and a worse RD performance for *CITIUSP* and *Facade9* PCs.

**Sparse PCs**

- Compared to the conventional MPEG standards, JPEG PCC outperforms G-PCC Octree and V-PCC Intra for all sparse PCs, except *Unicorn*. The gains in performance over G-PCC Octree come mostly from the lower rates, since at medium and higher rates G-PCC

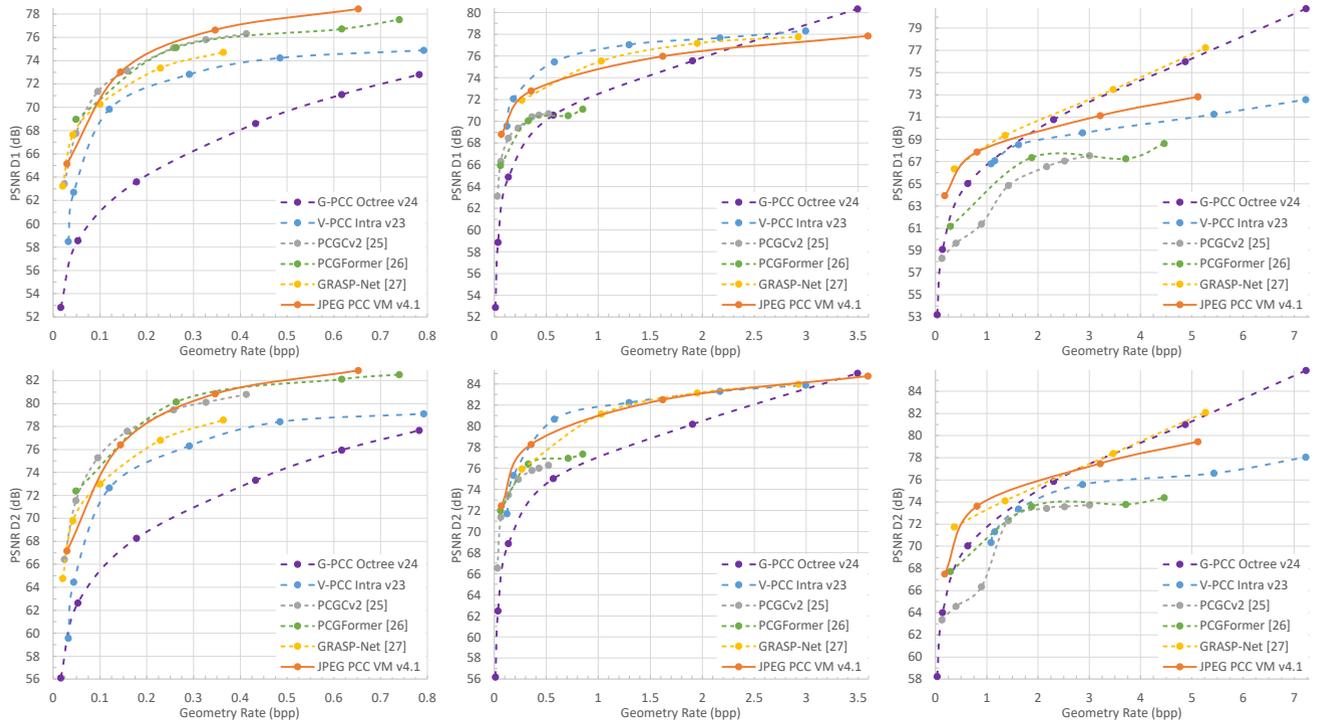

**FIGURE 10.** Average geometry-only RD performance of the JPEG PCC standard for the different categories' PCs (solid on the left, dense in the middle, and sparse on the right), in comparison with the benchmarks for the PSNR D1 (top) and PSNR D2 (bottom) quality metrics.

**TABLE 6.** BD-Rate and BD-PSNR performance of JPEG PCC in comparison with the benchmarks for PSNR D1 and PSNR D2.

| | Point Cloud | Ref: G-PCC Octree v24 | | | | Ref: V-PCC Intra v23 | | | | Ref: PCGCv2 [25] | | | | Ref: PCGFormer [26] | | | | Ref: GRASP-Net [27] | | | |
|---|---|---|---|---|---|---|---|---|---|---|---|---|---|---|---|---|---|---|---|---|---|
| | | PSNR D1 | | PSNR D2 | | PSNR D1 | | PSNR D2 | | PSNR D1 | | PSNR D2 | | PSNR D1 | | PSNR D2 | | PSNR D1 | | PSNR D2 | |
| | | BD Rate | BD PSNR | BD Rate | BD PSNR | BD Rate | BD PSNR | BD Rate | BD PSNR | BD Rate | BD PSNR | BD Rate | BD PSNR | BD Rate | BD PSNR | BD Rate | BD PSNR | BD Rate | BD PSNR | BD Rate | BD PSNR |
| **Solid** | *StMichael* | -85% | 8.2 | -75% | 6.8 | -60% | 4.1 | -57% | 4.8 | -16% | 0.7 | -14% | 0.7 | 8% | -0.1 | 1% | 0.2 | -24% | 1.2 | -28% | 1.7 |
| | *Bouquet* | -83% | 7.1 | -77% | 6.0 | -55% | 3.4 | -56% | 3.7 | 24% | -1.2 | 16% | -1.0 | 38% | -1.2 | 28% | -1.1 | -2% | 0.1 | -13% | 0.5 |
| | *Soldier* | -88% | 10.3 | -82% | 9.2 | -44% | 3.1 | -33% | 3.6 | -16% | 0.8 | -15% | 0.9 | 1% | 0.0 | 1% | 0.1 | -45% | 2.5 | -46% | 3.1 |
| | *Thaidancer* | -91% | 10.0 | -84% | 9.0 | -38% | 2.4 | -33% | 2.6 | -61% | 3.6 | -18% | 1.0 | -68% | 3.5 | 9% | -0.1 | -51% | 2.9 | -50% | 3.7 |
| **Dense** | *Boxer* | -80% | 5.7 | -89% | 8.5 | -41% | 1.0 | -40% | 1.9 | -79% | 3.2 | -65% | 4.3 | -91% | 3.2 | -31% | 3.1 | -42% | 0.5 | -47% | 1.5 |
| | *HouseWoRoof* | -58% | 3.1 | -61% | 3.7 | -25% | 0.5 | -59% | 2.5 | -52% | 1.4 | -61% | 1.9 | -73% | 2.1 | -55% | 1.7 | -8% | 0.1 | -33% | 1.0 |
| | *CITIUSP* | -32% | 1.5 | -20% | 0.9 | 3% | -0.1 | 8% | -0.3 | -65% | 3.2 | 20% | -0.3 | -61% | 3.7 | 35% | -0.3 | 27% | -0.9 | 29% | -1.1 |
| | *Facade9* | -31% | 1.8 | -54% | 3.5 | 580% | -5.7 | 132% | -3.4 | -46% | 1.5 | -45% | 1.8 | -63% | 2.0 | -37% | 1.6 | 9% | -0.3 | -32% | 1.1 |
| **Sparse** | *EPFL* | -28% | 1.0 | -15% | 0.4 | -23% | 0.7 | -13% | 0.4 | -4% | 0.5 | -15% | 1.1 | -61% | 2.6 | -37% | 1.3 | 13% | -0.5 | -10% | 0.1 |
| | *ArcoValentino* | -29% | 1.1 | -19% | 0.4 | -69% | 3.1 | -76% | 4.0 | -57% | 2.7 | -50% | 2.5 | -60% | 2.9 | -53% | 2.6 | -4% | -0.6 | -3% | -0.5 |
| | *Shiva* | -35% | 1.6 | -27% | 1.3 | -72% | 2.9 | -86% | 5.2 | -22% | 0.7 | -13% | 0.8 | -19% | 0.6 | 25% | -0.4 | 0% | -0.1 | -9% | 0.2 |
| | *Unicorn* | 45% | -1.6 | -39% | 1.8 | 37% | -1.6 | -25% | -1.7 | -94% | 15.2 | | 18.2 | -87% | 15.9 | -85% | 4.8 | 110% | -1.9 | 10% | 0.3 |
| | Average Solid | -87% | 8.9 | -80% | 7.8 | -49% | 3.2 | -47% | 3.7 | -17% | 1.0 | -8% | 0.4 | -5% | 0.6 | 10% | -0.2 | -30% | 1.7 | -34% | 2.3 |
| | Average Dense | -50% | 3.0 | -56% | 4.2 | 129% | -1.1 | 10% | 0.2 | -60% | 2.3 | -38% | 1.9 | -72% | 2.8 | -22% | 1.6 | -4% | -0.1 | -21% | 0.6 |
| | Average Sparse | -12% | 0.5 | -25% | 1.0 | -32% | 1.3 | -50% | 2.0 | -44% | 5.5 | -41% | 5.1 | -59% | 3.0 | -37% | 2.1 | 30% | -0.8 | -3% | 0.0 |

Octree can reach higher qualities. On the other hand, V-PCC Intra tends to underperform both G-PCC Octree and JPEG PCC quite significantly, with the exception of the *Unicorn* PC, for which V-PCC is able to outperform JPEG PCC. This shows that while V-PCC excels for solid and dense PC content, G-PCC is more appropriate for sparse PC content.

- Compared to the DL-based benchmarks, once again JPEG PCC significantly outperforms both PCGCv2 and

PCGFormer for all sparse PCs, with an average BD-Rate of -44% (-41%) and -59% (-37%), respectively, for PSNR D1 (PSNR D2). However, on average, JPEG PCC fails to outperform GRASP-Net; while reaching similar RD performance for low and medium rates, for high rates JPEG PCC tends to saturate the quality much faster, whereas GRASP-Net is able to continue increasing quality akin to G-PCC Octree.

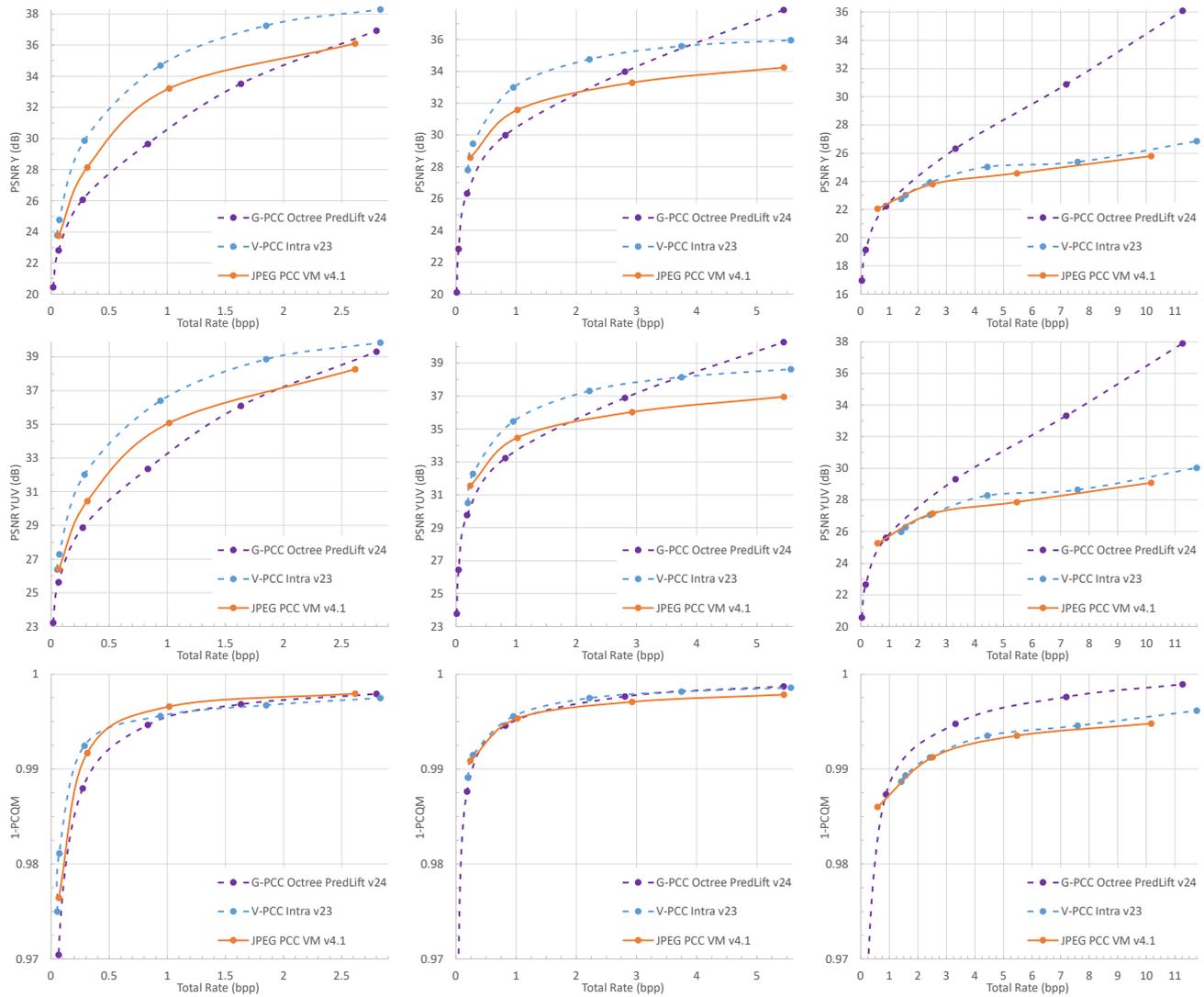

**FIGURE 11.** Average geometry and color RD performance of the JPEG PCC standard for the different categories' PCs (solid on the left, dense in the middle, and sparse on the right), in comparison with the benchmarks for the PSNR Y (top), PSNR YUV (middle), and 1-PCQM (bottom) metrics.

Overall, JPEG PCC offers a good PC geometry compression performance compared to the state-of-the-art, for both PSNR D1 and PSNR D2 metrics alike. Overall, JPEG PCC consistently achieves one of the best, if not the best, RD performance for all PC categories (solid, dense and sparse), although not necessarily being always the best.

### C. JOINT GEOMETRY AND COLOR RD PERFORMANCE

This sub-section assesses the JPEG PCC RD performance considering both geometry and color coding. Following the previously described experimental setup, the obtained results are presented in Fig. 11, which shows the RD charts with the average codec performance for each PC category (solid, dense, and sparse), for three different quality metrics: PSNR Y and PSNR YUV (luminance and chrominances), and the joint geometry and color 1-PCQM. Moreover, Table 7 details the BD-Rate and BD-PSNR results for JPEG PCC, taking the conventional MPEG standards benchmarks as reference, since the benchmark DL-based codecs are only able to encode the PC geometry (and not the color); results in bold represent the cases when JPEG PCC outperforms the benchmark. These results allow the following observations:

**Solid PCs**

- Compared to G-PCC Octree PredLift, JPEG PCC offers, on average, a better RD performance for both color metrics and more significantly for the joint metric, reaching average BD-Rate of -17%, -2%, and -27% for PSNR Y, PSNR YUV, and 1-PCQM, respectively. However, most of these gains come from the *Soldier* and *Thaidancer* PCs, where JPEG PCC outperforms consistently across all rates and all metrics; JPEG PCC fails to outperform G-PCC Octree PredLift for the *StMichael* and *Bouquet* PCs for the color metrics, mostly at higher rates where JPEG PCC saturates quickly.

**TABLE 7.** BD-Rate and BD-PSNR performance of JPEG PCC in comparison with the benchmarks for PSNR Y, PSNR YUV, and 1-PCQM.

| | Point Cloud | Ref: G-PCC Octree PredLift v24 | | | | | | Ref: V-PCC Intra v23 | | | | | |
|---|---|---|---|---|---|---|---|---|---|---|---|---|---|
| | | PSNR Y | | PSNR YUV | | 1-PCQM | | PSNR Y | | PSNR YUV | | 1-PCQM | |
| | | BD Rate | BD PSNR | BD Rate | BD PSNR | BD Rate | BD PSNR | BD Rate | BD PSNR | BD Rate | BD PSNR | BD Rate | BD PSNR |
| **Solid** | *StMichael* | 3% | -0.3 | 12% | -0.5 | **-12%** | **$9.8\times10^{-4}$** | 85% | -1.8 | 70% | -1.4 | 38% | $-2.8\times10^{-3}$ |
| | *Bouquet* | 16% | -0.6 | 45% | -1.1 | **-7%** | **$1.6\times10^{-6}$** | 113% | -2.0 | 124% | -1.9 | 26% | $-2.0\times10^{-3}$ |
| | *Soldier* | **-36%** | **2.1** | **-30%** | **1.5** | **-46%** | **$3.4\times10^{-3}$** | 59% | -2.3 | 50% | -1.8 | 1% | **$1.7\times10^{-4}$** |
| | *Thaidancer* | **-49%** | **2.9** | **-36%** | **1.9** | **-42%** | **$1.3\times10^{-3}$** | 41% | -1.7 | 48% | -1.8 | **-29%** | **$5.1\times10^{-4}$** |
| **Dense** | *Boxer* | **-53%** | **2.2** | **-38%** | **1.3** | **-58%** | **$2.0\times10^{-3}$** | 37% | -1.0 | 40% | -1.0 | **-6%** | **$2.0\times10^{-4}$** |
| | *HouseWoRoof* | 21% | -0.7 | 18% | -0.6 | 4% | $-1.2\times10^{-4}$ | 149% | -1.6 | 132% | -1.4 | 9% | $-2.5\times10^{-4}$ |
| | *CITIUSP* | **-8%** | 0.0 | 22% | -0.5 | 62% | $-1.5\times10^{-3}$ | 43% | -0.9 | 26% | -0.6 | 6% | $-2.9\times10^{-4}$ |
| | *Facade9* | **-32%** | **0.2** | **-36%** | **0.4** | 10% | $-3.5\times10^{-4}$ | 1% | -0.2 | **-17%** | **0.2** | 25% | $-5.0\times10^{-4}$ |
| **Sparse** | *EPFL* | 11% | -0.6 | 29% | -0.7 | 16% | $-5.7\times10^{-4}$ | **-27%** | **0.4** | **-21%** | **0.3** | 20% | $-4.6\times10^{-4}$ |
| | *ArcoValentino* | 64% | -1.7 | 48% | -1.3 | 11% | $-1.4\times10^{-3}$ | **-16%** | **0.2** | **-27%** | **0.3** | **-22%** | **$9.4\times10^{-4}$** |
| | *Shiva* | 62% | -2.6 | 60% | -2.2 | 78% | $-2.6\times10^{-3}$ | 44% | -0.6 | 65% | -0.6 | 8% | $-2.0\times10^{-4}$ |
| | *Unicorn* | 169% | -0.6 | 135% | -3.9 | 46% | $-1.1\times10^{-3}$ | 29% | **0.4** | 13% | -0.5 | 3% | $-3.0\times10^{-4}$ |
| | Average Solid | **-17%** | **1.0** | **-2%** | **0.4** | **-27%** | **$1.4\times10^{-3}$** | 75% | -1.9 | 73% | -1.7 | 9% | $-1.1\times10^{-3}$ |
| | Average Dense | **-18%** | **0.4** | **-8%** | **0.2** | 5% | **$1.9\times10^{-5}$** | 58% | -0.9 | 45% | -0.7 | 9% | $-2.1\times10^{-4}$ |
| | Average Sparse | 77% | -1.4 | 68% | -2.0 | 38% | $-1.4\times10^{-3}$ | 7% | **0.1** | 7% | -0.1 | 2% | $-5.1\times10^{-6}$ |

- Compared to V-PCC Intra, JPEG PCC considerably underperforms, especially for the color metrics. For the 1-PCQM joint metric, JPEG PCC seems to present a better RD performance for the medium and high rates, leading to a closer result of 9% BD-Rate, on average.

**Dense PCs**

- Compared to G-PCC Octree PredLift, JPEG PCC offers, on average, a slightly better RD performance for both color metrics, noticeable mostly in the low to medium rates, reaching an average BD-Rate of -18% and -8% for the PSNR Y and PSNR YUV metrics. However, for 1-PCQM, JPEG PCC's results are slightly worse on average, with a 5% BD-Rate increase, only outperforming G-PCC Octree PredLift for the *Boxer* PC.
- Compared to V-PCC Intra, once again JPEG PCC underperforms for all three metrics, with the closer results being for the joint metric 1-PCQM, showing a 9% BD-Rate.

**Sparse PCs**

- Compared to G-PCC Octree PredLift, JPEG PCC consistently underperforms for all three metrics. As mentioned before for the geometry-only compression performance, JPEG PCC is weakest for sparse PC content, whereas G-PCC Octree PredLift is more suited and performs better for this type of content.
- Compared to V-PCC Intra, JPEG PCC is still underperforming, although by a much smaller margin, as V-PCC Intra also tends to struggle with sparse PC content. Since JPEG PCC adopts the same projection approach as V-PCC Intra for color coding, it naturally inherits V-PCC's difficulties in coding this type of content, with both showing similar results and a RD

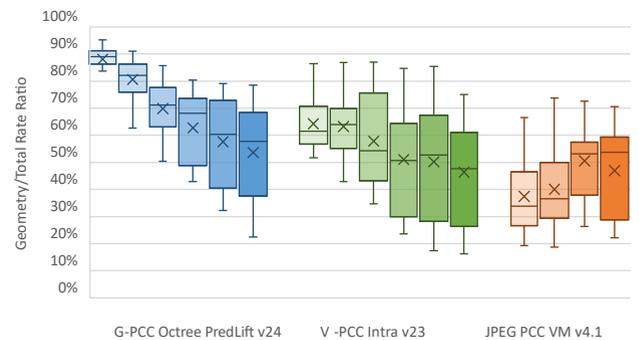

**FIGURE 12.** Geometry rate versus total rate ratio for the full test dataset considering each rate point (increasing from left to right) and each codec.

performance which is significantly below that of G-PCC Octree PredLift.

Overall, the JPEG PCC RD performance for PC color coding is not competitive with conventional MPEG standards. For solid and dense PCs in particular, V-PCC Intra shows considerably better RD performance for PSNR Y and PSNR YUV, despite JPEG PCC adopting the same projection approach. To get a better understanding of the RD performance, it is worth to analyze the allocation of rate for geometry and color in the full bitstream, for the selected JPEG PCC coding configurations and the G-PCC Octree PredLift and V-PCC Intra benchmarks. Fig. 12 shows the ratio between the geometry and total rates, for each rate point and each codec. G-PCC Octree PredLift shows a very clear trend where the large majority of the total rate (near 90%) is allocated to the geometry for the lowest rate point, with this ratio steadily decreasing as the total rate increases. V-PCC Intra shows a similar trend, although not as intense, with the three highest rate points showing a more similar ratio. On the other hand, JPEG PCC shows a somewhat opposite trend, with a smaller ratio at lower rate points, increasing slightly for the highest rate points. Overall, the geometry rate ratio also seems generally lower for JPEG PCC than for the G-PCC Octree PredLift and V-PCC Intra benchmarks. Contrary to what could be expected, the JPEG PCC color RD performance is inferior despite having a higher geometry rate ratio than G-PCC Octree PredLift and V-PCC Intra.

**TABLE 8.** Encoding and decoding times (in seconds) for JPEG PCC and the benchmarks for geometry-only and for geometry and color coding. Grey shading indicates the codecs ran on CPU and no shading indicates codecs ran partly on GPU.

| Point Cloud | | Geometry-only | | | | | | | | | | | | | | | | Geometry and Color | | | | | |
|---|---|---|---|---|---|---|---|---|---|---|---|---|---|---|---|---|---|---|---|---|---|---|---|
| | | G-PCC Octree v24 | | V-PCC Intra v23 | | PCGCv2 [25] | | PCGFormer [26] | | GRASP-Net [27] | | JPEG PCC VM v4.1 | | G-PCC Octree PredLift v24 | | V-PCC Intra v23 | | JPEG PCC VM v4.1 | |
| | | Enc | Dec | Enc | Dec | Enc | Dec | Enc | Dec | Enc | Dec | Enc | Dec | Enc | Dec | Enc | Dec | Enc | Dec |
| **Solid** | *StMichael* | 2.5 | 0.4 | 238 | 9.8 | 4.1 | 3.9 | 4.6 | 5.0 | 7.2 | 4.3 | 59.3 | 3.9 | 13.0 | 1.8 | 474 | 10.8 | 90.6 | 71.0 |
| | *Bouquet* | 4.6 | 0.8 | 370 | 16.8 | 4.3 | 4.5 | 6.8 | 7.1 | 12.4 | 4.7 | 141 | 7.6 | 22.3 | 3.2 | 943 | 18.8 | 196 | 108 |
| | *Soldier* | 1.2 | 0.3 | 74.6 | 4.0 | 4.0 | 3.9 | 4.0 | 4.1 | 5.0 | 4.2 | 22.1 | 2.9 | 7.0 | 1.1 | 182 | 4.6 | 44.0 | 36.2 |
| | *Thaidancer* | 2.2 | 0.3 | 184 | 13.6 | 8.3 | 8.7 | 11.9 | 12.2 | 14.8 | 9.3 | 97.4 | 8.8 | 25.1 | 1.3 | 504 | 16.0 | 145 | 99.5 |
| **Dense** | *Boxer* | 4.1 | 1.0 | 571 | 35.9 | 8.2 | 8.2 | 9.6 | 9.6 | 35.9 | 13.9 | 159 | 26.2 | 23.8 | 3.4 | 1154 | 43.0 | 276 | 278 |
| | *HouseWoRoof* | 8.6 | 1.6 | 1026 | 61.4 | 7.2 | 7.6 | 11.1 | 11.7 | 63.6 | 16.5 | 325 | 90.4 | 39.9 | 5.1 | 2226 | 74.1 | 673 | 689 |
| | *CITIUSP* | 5.9 | 1.3 | 1542 | 112 | 346 | 597 | 15802 | 15926 | 77.3 | 26.0 | 278 | 56.2 | 52.8 | 5.6 | 2965 | 137 | 553 | 395 |
| | *Facade9* | 3.3 | 0.8 | 635 | 29.7 | 7.1 | 7.2 | 8.2 | 8.4 | 18.7 | 11.0 | 123 | 39.8 | 12.7 | 2.2 | 1116 | 37.5 | 294 | 351 |
| **Sparse** | *EPFL* | 6.2 | 1.5 | 2495 | 162 | 393 | 745 | 25971 | 26145 | 60.8 | 22.0 | 179 | 36.6 | 34.8 | 4.5 | 5279 | 196 | 553 | 284 |
| | *ArcoValentino* | 3.9 | 1.1 | 1257 | 51.2 | 7.5 | 7.9 | 10.6 | 11.5 | 18.8 | 10.6 | 96.4 | 27.8 | 13.8 | 2.8 | 1794 | 61.9 | 307 | 189 |
| | *Shiva* | 2.5 | 0.8 | 542 | 35.3 | 7.0 | 7.5 | 9.1 | 9.3 | 13.4 | 9.5 | 102 | 36.9 | 9.1 | 1.9 | 1406 | 49.6 | 366 | 291 |
| | *Unicorn* | 4.7 | 1.4 | 1976 | 198 | 288 | 1193 | 22151 | 22723 | 28.5 | 19.4 | 171 | 73.2 | 17.4 | 3.4 | 3940 | 242 | 836 | 587 |

### D. COMPUTATIONAL COMPLEXITY

In addition to the RD performance, it is also important to assess the JPEG PCC computational complexity in comparison to the benchmarks. In the literature, it is common to measure the complexity of DL models in terms of floating-point operations per second (FLOPS) or multiply–accumulate operations (MAC). However, due to the nature of sparse convolutions, which were implemented with the popular Minkowski Engine library [50], it was not possible to compute such complexity metrics. Furthermore, since in sparse convolutions the kernel is only applied to the occupied voxels, the number of operations is highly dependent both on the number of points and its distributions, i.e. on the PC content itself. In this context, the computational complexity was measured in terms of the average encoding and decoding times (in seconds) for each PC in the test dataset, using an Intel Core i7-14700k CPU @ 5.60 GHz computer with a NVidia GeForce RTX 4090 24 GB GPU and 64 GB of RAM, running Debian 12. This allows a relative performance analysis since all codecs were run under the same computational platform and conditions.

Table 8 shows the complete results for JPEG PCC and all benchmarks, considering both geometry-only coding and geometry and color coding. The MPEG standards were run on CPU (shown with a grey shading) whereas the DL-based benchmarks were run partly on GPU (shown without shading), with the exception of the 13-bit precision PCs for the PCGCv2 and PCGFormer benchmarks which were not able to run on GPU due to memory issues; in these cases, the encoding and decoding times are significantly higher and comparisons do not apply. For JPEG PCC, the DL coding and SR models as well as JPEG AI were run on GPU, whereas modules such as the binarization optimizations, recoloring, 2D to 3D projection, 3D to 2D inverse projection, and color super-resolution were run on CPU. Because of the mix of very different CPU and GPU results, no averages are presented. It is also important to refer that these computational complexity results have only a limited qualitative value since the used software have limited or no optimization, as it is the case for the used JPEG PCC reference software; optimization would significantly help to reduce the encoding and decoding times.

The results for geometry-only coding show that G-PCC Octree is the least computationally complex codec, requiring only a few seconds to encode and even less to decode all PCs, even though running on CPU. On the opposite end, V-PCC Intra is clearly the most complex, especially for encoding, taking more than 1000 seconds for some PCs. As for the DL-based benchmarks, in general, they tend to be closer to the G-PCC Octree encoding times, with PCGCv2 being the least complex, followed by PCGFormer and, finally, GRASP-Net. Regarding JPEG PCC, the encoder is considerably more complex than GRASP-Net, mostly due to the binarization optimizations module, though still very much below V-PCC Intra.

For geometry and color coding, a similar trend is observed, where G-PCC Octree PredLift is the least computationally complex codec and V-PCC Intra is the most computationally complex codec, with JPEG PCC placed in between. However, unlike for geometry-only coding, the decoding times for JPEG PCC are higher than V-PCC Intra, likely because JPEG PCC must perform the 3D to 2D projection again at the decoder to produce the information required for the inverse projection, as described in Section III.F.

### E. OVERALL ANALYSIS AND DISCUSSION

The JPEG PCC RD performance for geometry coding is very competitive against the state-of-the-art codecs, including the conventional MPEG standards and learning-based solutions in the literature. On the other hand, the JPEG PCC RD performance for color coding is less competitive, failing to outperform V-PCC Intra and G-PCC Octree PredLift in most cases. However, the added value of a learning-based coding standard does not only regard its RD performance since it offers other valuable functional benefits over existing standards; moreover, there is still a significant margin for RD performance improvement which may be exploited in the future, as highlighted in the following.

#### 1) COMPRESSED DOMAIN REPRESENTATIONS FOR MAN AND MACHINE

The key functional benefit of the JPEG PCC standard is its capability to provide a unified compressed representation efficiently and effectively serving both human visualization and machine processing; this is a breakthrough in visual representation never before possible with conventional coding. This is achieved with the adoption of DL coding models both for geometry and color coding, which both produce a compressed latent representation encoded into the bitstream. This way, the same bitstream can be processed either by a regular decoder, allowing to reconstruct the PC for human visualization, or directly by a computer vision processor for a machine vision task, without decoding. In addition to being conceptually elegant, this unified approach also provides advantages in terms of reduced computational complexity (no decoding is needed) and increased computer vision tasks performance (features are extracted from the original and not lossy decoded PC), as already demonstrated in practice [57][58]; this is a very attractive and powerful functionality which practical impact may now be exercised with the specification of JPEG PCC standard.

#### 2) SINGLE CODEC FOR ALL PC CONTENT

Another strength of JPEG PCC is that it adopts the same trained DL coding models for all types of PC content, such as solid, dense, and sparse, while still able to maintain a consistently high RD performance for geometry coding, which is not the case for other tested benchmarks. For example, MPEG has developed two PC coding standards targeting different PC content, with G-PCC specializing in static and sparser PCs and V-PCC Intra specializing in static solid/denser PCs; when applied to different PC content, both severely underperform, as is the case with G-PCC Octree for solid PCs and V-PCC Intra for sparse PCs. As for the DL-based codecs, PCGCv2 and PCGFormer have been developed targeting mostly solid PC content [25][26], thus their performance for dense and sparse PCs tends to suffer, whereas GRASP-Net [27] uses different DL models trained specifically for each PC category, what leads to a better RD performance for sparse content.

#### 3) GEOMETRY AND COLOR RATE ALLOCATION

JPEG PCC offers the freedom to independently select the geometry and color coding configurations for each PC, thus allowing to dynamically and adaptively define the geometry and color rate allocation in the bitstream. The RD performance assessed in this paper considers a specific JPEG PCC coding configuration, selected to optimize the joint quality measured with the 1-PCQM metric. However, this is not necessarily the best approach, and RD performance may be improved with different strategies for selecting the optimal geometry and color rate allocation, which are possible to deploy since they are non-normative.

Furthermore, the RD performance is highly dependent on the distortion/quality metrics used for selecting the optimal coding configuration and evaluation. Since the field of PC objective quality assessment is still fairly recent compared to image and video, it is expectable that better PC quality metrics emerge in the future, thus allowing achieving better RD performance. With this in mind, a study involving subjective quality assessment has been recently performed [59] to determine the best rate allocation between geometry and color for JPEG PCC, as well as G-PCC Octree PredLift and V-PCC Intra.

#### 4) JPEG AI TRAINING DATASET

Currently, JPEG PCC uses the JPEG AI coding models without any modification to achieve full compatibility. However, while JPEG AI was optimized and trained essentially with natural image content, the JPEG PCC projected images to be coded are substantially different from natural images, as shown in Fig. 6. Since training plays an extremely important role in any DL-based solution, this mismatch between the training content and the content that will actually be coded may make JPEG PCC color coding less efficient. One way to mitigate this mismatch would be to train new, adapted JPEG AI coding models using a training dataset consisting of projected images; these new models should have the potential to improve the RD performance. Since both JPEG PCC and JPEG AI are part of the JPEG ecosystem, it would eventually be possible to create a new JPEG AI profile for such content, which could be seamlessly integrated in the current JPEG PCC standard without any changes.

#### 5) JPEG AI TRAINING LOSS FUNCTION

In line with the previous discussion, the development and optimization of the JPEG AI coding models was heavily influenced by quality metrics which more closely correlate to the subjective visual quality, including MS-SSIM during training. However, the benefits of optimizing for these metrics may not translate directly to a better quality when using fidelity metrics such as PSNR Y and PSNR YUV for assessing the PC color coding performance; this contrasts with VVC which optimization mostly involves the MSE distortion metric, which may fit better the PSNR Y and PSNR YUV PC color quality metrics. This is also corroborated by the fact that JPEG PCC's results for the 1-PCQM metric tend to be better, approaching or even surpassing the conventional MPEG standards, since 1-PCQM is a PC quality metric better correlating with the subjective visual quality [56]. As previously mentioned, the creation of a new JPEG AI profile with coding models trained with a dataset of projected images may also adopt different loss functions, in particular considering one or more metrics that better correlate with the PC quality assessment metrics.

**TABLE 9. BD-Rate and BD-PSNR performance of JPEG PCC (with SR) in comparison with JPEG PCC without SR as reference, for PSNR D1 and PSNR D2.**

| | Point Cloud | PSNR D1 | | PSNR D2 | |
|---|---|---|---|---|---|
| | | BD Rate | BD PSNR | BD Rate | BD PSNR |
| **Solid** | StMichael | -50% | 3.3 | -38% | 2.5 |
| | Bouquet | -17% | 1.3 | -18% | 1.3 |
| | Soldier | -48% | 4.9 | -41% | 4.1 |
| | Thaidancer | -45% | 3.2 | -36% | 2.8 |
| **Dense** | Boxer | -71% | 2.1 | -73% | 4.6 |
| | HouseWoRoof | -22% | 0.6 | -31% | 1.4 |
| | CITIUSP | -1% | 0.1 | 0% | 0.1 |
| | Facade9 | 15% | -0.4 | -20% | 0.9 |
| **Sparse** | EPFL | 0% | 0.0 | 0% | 0.0 |
| | ArcoValentino | 0% | 0.0 | 0% | 0.0 |
| | Shiva | -6% | 0.2 | -7% | 0.3 |
| | Unicorn | 0% | 0.0 | 0% | 0.0 |
| | Average Solid | -40% | 3.2 | -33% | 2.7 |
| | Average Dense | -20% | 0.6 | -31% | 1.8 |
| | Average Sparse | -2% | 0.0 | -2% | 0.1 |

**TABLE 10. BD-Rate and BD-PSNR performance of JPEG PCC using VVC in comparison with JPEG PCC using JPEG AI as reference, for PSNR Y, PSNR YUV and 1-PCQM.**

| | Point Cloud | PSNR Y | | PSNR YUV | | 1-PCQM | |
|---|---|---|---|---|---|---|---|
| | | BD Rate | BD PSNR | BD Rate | BD PSNR | BD Rate | BD PSNR |
| **Solid** | StMichael | -22% | 0.7 | -31% | 0.9 | -6% | $5.5 \times 10^{-3}$ |
| | Bouquet | -21% | 0.6 | -35% | 1.0 | -5% | $1.8 \times 10^{-4}$ |
| | Soldier | -12% | 0.5 | -21% | 0.9 | 4% | $-2.0 \times 10^{-4}$ |
| | Thaidancer | -8% | 0.4 | -25% | 1.3 | 6% | $-8.4 \times 10^{-5}$ |
| **Dense** | Boxer | -8% | 0.0 | -33% | 0.6 | 6% | $-1.4 \times 10^{-4}$ |
| | HouseWoRoof | -16% | 0.3 | -32% | 0.6 | 14% | $-2.9 \times 10^{-4}$ |
| | CITIUSP | -30% | 0.5 | -44% | 0.8 | -14% | $3.6 \times 10^{-4}$ |
| | Facade9 | -14% | 0.3 | -30% | 0.7 | -10% | $1.5 \times 10^{-4}$ |
| **Sparse** | EPFL | -20% | 0.3 | -38% | 0.7 | -1% | $1.9 \times 10^{-5}$ |
| | ArcoValentino | -28% | 0.3 | -38% | 0.4 | -3% | $1.0 \times 10^{-4}$ |
| | Shiva | -24% | 0.3 | -43% | 0.7 | 8% | $-2.3 \times 10^{-4}$ |
| | Unicorn | -2% | 0.3 | -29% | 0.6 | 6% | $-6.9 \times 10^{-5}$ |
| | Average Solid | -16% | 0.6 | -28% | 1.0 | 0% | $1.1 \times 10^{-4}$ |
| | Average Dense | -17% | 0.3 | -35% | 0.7 | -1% | $2.1 \times 10^{-5}$ |
| | Average Sparse | -18% | 0.3 | -37% | 0.6 | 2% | $-4.4 \times 10^{-5}$ |

## V. ABLATION AND REPLACEMENT ANALYSIS

This section intends to demonstrate the importance of some modules and evaluate their impact on the RD performance of JPEG PCC by disabling or replacing them.

### A. SUPER-RESOLUTION

This experiment assesses the impact of using SR in the PC geometry coding RD performance. For this purpose, all PCs in the test dataset were coded following the same coding configurations in Table 5, keeping the down/up-sampling with the chosen SF, but disabling the DL-based geometry SR module, described in Section III.C.

The obtained results are presented in Table 9, containing the BD-Rate and BD-PSNR performance for the JPEG PCC standard (with SR) in comparison with the reference JPEG PCC without SR; results in bold represent the cases when JPEG PCC with SR outperforms the reference JPEG PCC without SR. From these results, it is clear that SR has a very significant impact on the geometry RD performance, in particular for solid and dense PCs, resulting in an average BD-Rate of -40% (-33%) and -20% (-31%) for solid and dense PCs, respectively, for PSNR D1 (PSNR D2) metric. On the other hand, the impact for sparse PCs is not significant as SR is rarely used in the coding configuration since it does not provide a benefit in the reconstruction quality for this type of content.

### B. 2D IMAGE CODING

This experiment evaluates the impact of the 2D image codec used to code the projected color images. As mentioned in Section IV.E, the current JPEG AI standard may not be the most efficient solution for coding the type of images resulting from the PC projection, in part due to having been trained and optimized for natural images (and not projected patches). To understand the potential impact of this approach, this experiment replaces JPEG AI with VVC Intra coding as the 2D image codec used by JPEG PCC. The JPEG PCC coding configurations for the geometry were kept the same as detailed in Table 5, but now the VVC test model version 23.3 was used, following the coding configurations specified in [52]. For each PC and RD point, the best VVC quantization parameter was selected to approximate the rate achieved with JPEG AI, thus expecting quality gains.

The obtained results are presented in Table 10, containing the BD-Rate and BD-PSNR values for JPEG PCC with VVC Intra in comparison with the reference JPEG PCC standard (with JPEG AI); results in bold represent the cases when using JPEG PCC with VVC Intra outperform the reference JPEG PCC with JPEG AI. The results demonstrate that using VVC Intra as the image codec provides a significantly better RD performance for both color metrics, consistently for all PCs and all PC categories: average BD-Rate between -16% and -18% for PSNR Y and between 28% and 37% for PSNR YUV are obtained.

While during the development of the JPEG AI standard it has been reported that it outperforms VVC Intra by as much as 30% BD-Rate for natural images [22][48], it is important to highlight that these results do not consider the PSNR Y, PSNR U and PSNR V image quality metrics, which are more widely adopted in conventional coding standardization activities. These are fidelity metrics which compute the distortion purely as a mathematical error and have been found to not have a very high correlation with subjective assessment [60]; this has pushed the JPEG AI standardization activity to adopt seven other image quality metrics which better correlate with the subjective assessment. As such, the JPEG AI standard was developed and optimized taking into account more than usual the subjective visual quality. This is evidenced by the results for the 1-PCQM joint metric, which is more correlated with the subjective visual quality [56], shown in Table 10. In this case, the difference between using JPEG AI and using VVC Intra is not as significant, with VVC Intra even slightly losing for some PCs.

While the improvements demonstrated in this experiment may not be enough to completely outperform the conventional MPEG standards G-PCC Octree PredLift and V-PCC Intra, they certainly help bridging the gap, and ultimately demonstrate the potential of the JPEG PCC standard. With the possibility of training JPEG AI's DL models for projected PC image content, eventually creating a new JPEG AI profile, there is a large margin for improvement on the JPEG PCC RD performance for color coding. Just substituting JPEG AI with VVC Intra is not a promising solution since this would break the JPEG PCC learning-based coding framework involving geometry and color. Moreover, it is well known that VVC Intra is heavily burdened by licensing, a dimension that is still unknow for learning-based coding.

## VI. CONCLUSION

This paper offers a detailed description of the JPEG Pleno Learning-based Point Cloud Coding standard, which aims at providing a unified compressed representation to serve both man and machine. Together with JPEG AI for image coding, these are the first two standards of a new coding generation exploiting data-based learning, which impact may start to emerge. While JPEG PCC RD performance still has margin for improvement in the next years, notably exploiting the methods already referred in this paper, the standard in its current stage 1 already offers a novel powerful type of representation able to effectively unify PC compressed domain for man and machine. In future work, the planned development of stages 2 and 3 will begin, with the goal of adding the support for enhanced reconstruction for human visualization and computer vision tasks from the same compressed domain representation, respectively. In addition, stage 2 of JPEG PCC may also continue to improve its RD performance, introducing novel tools (general or content specific) such as attention models. Furthermore, the requirements not yet addressed in the current version of JPEG PCC will be considered, such as quality scalability and lossless coding.

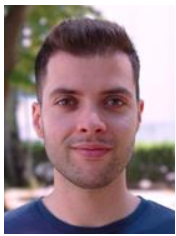

**ANDRÉ F. R. GUARDA** (Member, IEEE) received his B.Sc. and M.Sc. degrees in electrotechnical engineering from Instituto Politécnico de Leiria, Portugal, in 2013 and 2016, respectively, and the Ph.D. degree in electrical and computer engineering from Instituto Superior Técnico, Universidade de Lisboa, Portugal, in 2021. He has been a researcher at Instituto de Telecomunicações since 2011, where he currently holds a Post-Doctoral position. His main research interests include multimedia signal processing and coding, with particular focus on point cloud coding with deep learning. He has authored several publications in top conferences and journals in this field, and is actively contributing to the standardization efforts of JPEG and MPEG on learning-based point cloud coding.

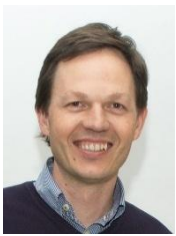

**NUNO M. M. RODRIGUES** (Senior Member, IEEE) graduated in electrical engineering in 1997, received the M.Sc. degree from the Universidade de Coimbra, Portugal, in 2000, and the Ph.D. degree from the Universidade de Coimbra, Portugal, in 2009, in collaboration with the Universidade Federal do Rio de Janeiro, Brazil. He is a Professor in the Department of Electrical Engineering, in the School of Technology and Management of the Polytechnic University of Leiria, Portugal and a Senior Researcher in Instituto de Telecomunicações, Portugal. He has coordinated and participated as a researcher in various national and international funded projects. He has supervised three concluded PhD theses and several MSc theses. He is co-author of a book and more than 100 publications, including book chapters and papers in national and international journals and conferences. His research interests include several topics related with image and video coding and processing, for different signal modalities and applications. His current research is focused on deep learning-based techniques for point cloud coding and processing.

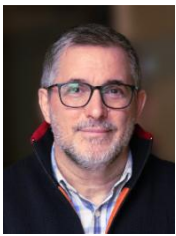

**FERNANDO PEREIRA** (Fellow, IEEE) graduated in electrical and computer engineering in 1985 and received the M.Sc. and Ph.D. degrees in 1988 and 1991, respectively, from Instituto Superior Técnico, Technical University of Lisbon. He is with the Department of Electrical and Computers Engineering of Instituto Superior Técnico, University of Lisbon, and Instituto de Telecomunicações, Lisbon, Portugal. He is or has been Associate Editor of IEEE Transactions of Circuits and Systems for Video Technology, IEEE Transactions on Image Processing, IEEE Transactions on Multimedia, IEEE Signal Processing Magazine and EURASIP Journal on Image and Video Processing, and Area Editor of the Signal Processing: Image Communication Journal. In 2013-2015, he was the Editor-in-Chief of the IEEE Journal of Selected Topics in Signal Processing. He was an IEEE Distinguished Lecturer in 2005 and elected as an IEEE Fellow in 2008 for "contributions to object-based digital video representation technologies and standards". He has been elected to serve on the IEEE Signal Processing Society Board of Governors in the capacity of Member-at-Large for 2012 and 2014-2016 terms. He has been IEEE Signal Processing Society Vice-President for Conferences in 2018-2020 and IEEE Signal Processing Society Awards Board Member in 2017. He was the recipient of the 2023 Leo L. Beranek Meritorious Service Award. Since 2013, he is also a EURASIP Fellow for "contributions to digital video representation technologies and standards". He has been elected to serve on the European Signal Processing Society Board of Directors for a 2015-2018 term. He was the recipient of the 2023 EURASIP Meritorious Service Award. Since 2015, he is also an IET Fellow. He has also held key leadership roles in numerous IEEE Signal Processing Society conferences and workshops, mostly notably serving twice as ICIP Technical Chair in two continents, Hong Kong (2010) and Phoenix (2016). He has been MPEG Requirements Subgroup Chair and is currently JPEG Requirements Subgroup Chair. Recently, he has been one of the key designers of the JPEG Pleno and JPEG AI standardization projects. He has contributed more than 300 papers in international journals, conferences and workshops, and made several tens of invited talks and tutorials at conferences and workshops. His areas of interest are video analysis, representation, coding, description and adaptation, and advanced multimedia services.